\documentclass[10pt]{article}
\usepackage{geometry}                		
\usepackage{graphicx}
\usepackage{multicol}        
\usepackage{hyperref}
\usepackage{amssymb}
\usepackage{amsmath,color,xcolor}
\usepackage{array}
\usepackage{graphics}
\usepackage{graphicx}
\usepackage{array}
\usepackage{fancyhdr}
\usepackage{enumerate}
\usepackage{verbatim}

\newcommand{\revFirst}[1]{{\leavevmode\color{black}#1}}

\usepackage{authblk}

\title{Neutrino telescopes and high-energy cosmic neutrinos} 

\author{Andrea Palladino$^{1}$,
Maurizio Spurio$^{2,3}$,
and Francesco Vissani$^{4,5,}$\\ \footnotesize
$^{1}$ Deutsches Elektronen-Synchrotron (DESY), Platanenallee 6, D-15738 Zeuthen, Germany\\ 
$^{2}$ NFN-Sezione di Bologna, Viale Berti-Pichat 6/2, 40127 Bologna, Italy\\
$^{3}$ Dipartimento di Fisica e Astronomia, dell'Universit\`a, Viale Berti Pichat 6/2, 40127 Bologna, Italy  
\\
$^{4}$  INFN, Laboratori Nazionali del Gran Sasso, 67100  L'Aquila, Italy\\
$^{5}$ Gran Sasso Science Institute, 67100 L'Aquila, Italy
} 

\date{}

\begin{document}

\maketitle



\begin{abstract}In this review paper, we present the main aspects of high-energy cosmic neutrino astrophysics. 
We begin by describing the generic expectations for cosmic neutrinos, including the effects of propagation from their sources to the detectors. 
Then we introduce the operating principles of current neutrino telescopes, and examine the main features  (topologies) of the observable events. After a discussion of the main background processes, due to the concomitant presence of secondary particles produced in the terrestrial atmosphere by cosmic rays, we summarize the current status of the observations {with astrophysical relevance} 
that have been greatly contributed by IceCube detector.
Then, we examine various  interpretations of these findings,  trying to assess the best candidate sources of cosmic neutrinos. 
We conclude with a brief perspective on how the field could evolve within a few years.
\end{abstract}

\parskip1ex

{
\section{Introduction}

\subsection{Historical~Notes}
The search for high-energy neutrinos from the cosmos is closely linked to cosmic rays and particle physics. In~fact, cosmic ray collisions produce particles  from whose decays the neutrinos are born, mainly  charged pions and muons. So, the~roots of high-energy neutrino astronomy are linked to the names of Hess, Pacini, Pauli, Fermi, Yukawa, Anderson, Neddermeyer, Powell, Lattes, Occhialini and other famous physicists of the early 20th~century.


History begins with a handful of theoretical works.
In the middle of the Cold War,  two independent papers~\cite{greisen, markov} proposed to  the scientific communities of the opponent countries the (same) idea, namely, of~building high-energy neutrino telescopes.
This is why the birth of this field of research is often attributed to these two works, but~earlier explorations had been conducted in a Diploma thesis presented in 1958 at Moscow State University by  Zheleznikh, by~the time a student of Markov: see~\cite{markov} and~\cite{zh2}. 
In this thesis it is written ``It is worth  searching 
 for high energy neutrinos from Outer Space, especially, if~the high energy $\gamma$-rays beyond the atmosphere were found'' \cite{zh2}: in other words,  a~very close connection between neutrinos and high-energy gamma rays was expected from the very beginning, 
because of the associated and inevitable production of charged and neutral pions.
Incidentally, the~study of gamma-ray radiation began in the same years by the paper of Morrison~\cite{morr} (who mentions ``the still unexploited neutrino channel'', even if only in connection to stellar processes)
and it is one of the most interesting branches of astronomy today.
We cite the inspired words from Reference~\cite{greisen}: ``one may predict that these fields of high-energy quantum and neutrino astronomy will be opened up in the near future'' which are particularly interesting, as~nowadays there is occasionally still some resistance 
in accepting the consistence of ``neutrino~astronomy''.

The first experimental searches for high energy neutrinos were conducted soon later 
by two experiments (KGF in India and CWI 
 in South Africa, which were attended by scientists from all over the world, 
 including the host countries, USA and USSR, UK, China and Japan)
who released their final results in the beginning of the 1970s~\cite{atm2,atm3,atm1}  and 
 revealed for the first time atmospheric neutrinos {but did not find evidence of  other components in their dataset. 
Shortly afterwards the idea emerged of building a much larger detector, about a kilometer in size, 
as the  way to proceed in the search for high-energy neutrinos from the cosmos.} 
This led to the proposal of one experiment named DUMAND, carried out by scientists from USA and USSR; however, the~well-known 
geopolitical circumstances prevented its realization. The~first detector that succeeded in this goal has been IceCube, which built upon the experience of AMANDA, 1996--2005. In~2013, IceCube showed the first evidence of a new component of cosmic neutrinos~\cite{disc}, whose nature, however, is not yet~identified. 

For further historical discussion, see Reference~\cite{syn,spie,storione}.
\subsection{Plan of This~Review}

The plan of this review paper is as follows: 
we begin by expounding the general expectations on cosmic neutrinos; next, we describe the principles of neutrino telescopes; 
we outline the possible signals and the background processes; then,  we summarize the main observational results {that have astrophysical 
significance}; finally, we consider a few specific astrophysical models, as~possible interpretations of these findings. In~fact, even though numerous theoretical proposals preceded the discovery, it does not seem fair to say that the theory provided a solid guide to the~discovery. 

However, before~proceeding in the discussion, it seems useful to 
conclude this brief introduction by highlighting a last momentous point, namely 
the discovery of the phenomenon of neutrino oscillations, anticipated in the decade 1957--1967
by Pontecorvo's theories~\cite{ponteggio} and fully accomplished thanks to a long series of experiments, 
including the Neutrino Detector Experiment in the Kamioka mine
(Super-KamiokaNDE) in Japan and the Sudbury Neutrino Observatory (SNO) in Canada 
 recognized by the Nobel prize in Physics in 2015~\cite{2015}.
As this aspect has direct implications on the field of cosmic neutrinos of very high energies, and~it seems to be one of the most reliable 
points in the theoretical interpretation, we will discuss in detail here.}

\section{High-Energy Neutrinos in the~Cosmos}

\subsection{The Cosmic Ray~Connection\label{sec:CRng}}
It is widely believed that certain cosmic sources produce high-energy neutrinos.  
The simplest reason is just the knowledge of {\em atmospheric neutrinos,} that originate from cosmic ray interactions with Earth's atmosphere. 
Many variants of the known mechanism, which could lead to a potentially observable population of {\em cosmic neutrinos,} can be imagined.  However, the~reference picture for the production of high energy neutrinos is simply that, the~same sites where cosmic rays are accelerated - or the environment that surround them---if endowed with sufficient target to convert a fraction of the energy into secondary particles, are potentially observable sources of high energy~neutrinos.

Therefore, neutrino astronomy could be ultimately a major avenue to discover the sources of cosmic rays, which are still 
{a} matter of theoretical speculation and to date not yet sufficiently known. 
For this reason, we begin by collecting the main important facts on cosmic~rays.

It is useful to recall some important observational facts concerning cosmic rays. 
The observation of high-energy part of cosmic ray spectrum at Earth allows us to distinguish several pieces, continuously connected but different among them. 
The flux is characterized by four main energy intervals: 
(1) the part below the {\em knee} at about $E_{\mbox{\tiny knee}}\sim 3\mbox{ PeV}$; (2)~the part just above, till~the feature called the {\em ankle,} which is at about {$E_{\mbox{\tiny ankle}}\sim 5\mbox{ EeV}$}; (3)~the {subsequent} decade of energies; 4)~the {(GZK) $suppression$ region, which corresponds to} the end of the observed spectrum. 
In all these regions, except~the last one the distribution can be roughly described as \textbf{power laws,} namely by energy distributions of the form:
\begin{equation}
\Phi_{\mbox{\tiny CR}}\sim E_{\mbox{\tiny CR}}^{-\alpha}\ ,
\label{eq:cr1}
\end{equation}
which are the characteristic features of the occurrence of (highly) non-thermal processes. 
The parameter $\alpha$ is called \textit{the slope.} 
The observed slopes {in the different energy intervals} are not directly related to the slope at production; effects of propagation and/or of escape are plausibly also energy dependent. Indeed, the~mechanisms of cosmic ray accelerations---for example, the~Fermi mechanism---are expected to produce in several cases a distribution with slope $\alpha\sim 2$ until the maximum energy they can reach, which depends upon the~accelerator. 

A few general references are helpful for deepening the study including References~\cite{astro-book-m,astro-book} and the reviews in References~\cite{a1,a4,a5,a3, a6,a7,a8,epj+}. 
For a discussion of the very important link with cosmic rays (briefly examined in this section) see References
~\cite{a4,baur,ger,a2,matthiae}. 

\subsection{The $pp$ and $p\gamma$ Mechanisms of Neutrino~Production\label{sec:hm}}
We assume that (most of) the observable cosmic neutrinos derive from astrophysical mechanism, that is, from~cosmic ray collisions. Consider the case of very high energy protons (or nucleons) colliding over a nucleus $\mathcal{N}$ at rest, described approximately as an assembly of nucleons.
This collision will lead to the production of many $\pi^+,\pi^-,\pi^0$, roughly in similar amount that we can resume symbolically by:
\begin{equation}
p+\mathcal{N}\to X+\mbox{many}\times ( \pi^+ + \pi^- + \pi^0).
\label{eq:pp1}
\end{equation}

The experimental observation shows that the highest energy pion (called the leading pion) 
carries on average 1/5 of the initial kinetic energy of the proton---see for example, Reference~\cite{berez,leading}. 
Its decay will eventually lead to high-energy neutrinos and gamma rays:
\begin{equation}
\pi^0\to \gamma+\gamma \ \mbox{ and} \ \pi^\pm\to \mu^\pm + \stackrel{(-)}\nu_\mu\ ,\mbox{ followed e.g.,~by }
\mu^+\to e^++\nu_e+\bar{\nu}_\mu.
\label{eq:pp2}
\end{equation}

The kinematic of these decays is such that, in~the decay of the neutral  $\pi^0$, each gamma rays carries 1/2 of the initial energy, and~likewise, in~the decay chain of the charged leptons, anyone of the four light (anti)leptons carries away 1/4 of the initial energy 
{(this corresponds to the result of the accurate calculation~\cite{volkova}; see also Appendix~\ref{appa}).} 
Thus, we have the very simple and useful \revFirst{rule of thumb}:
\begin{equation}\label{eq:pp3}
E_{\nu} \sim \frac{E_\gamma}{2}\sim \frac{E_\pi}{4} \sim \frac{E_p}{20} \ .
\end{equation}

Interestingly, this remains unchanged also for a completely different case, when the target is replaced by (high-energy) gamma rays.  In~fact, consider the process:
\begin{equation}\label{eq:pg1}
p+\gamma\to \Delta^+\to n+ \pi^+ \mbox{ or alternatively } \to p+ \pi^0.   
\end{equation}

By simple kinematic considerations, one obtains that, on~average, 
the mesons carry away 1/5 of the initial kinetic energy; therefore, the~same \revFirst{rule of thumb} (\ref{eq:pp3}) applies. For~instance:
\revFirst{$$
E_\nu =100\mbox{ TeV}-10 \mbox{ PeV} \mbox{ corresponds to }E_p\sim 2\mbox{ PeV}-200, \mbox{ PeV}
$$ }
which is the region above (or just around) the knee of the cosmic rays spectrum.
Note that no electron antineutrinos are formed in the decay chain of Equation~(\ref{eq:pg1}). 
These appear after oscillations (see Section~\ref{sec:osci}) even if a relative shortage of $\bar\nu_e$ persists.

\begin{figure}[t]
\begin{center}
\includegraphics[width=0.8\textwidth]{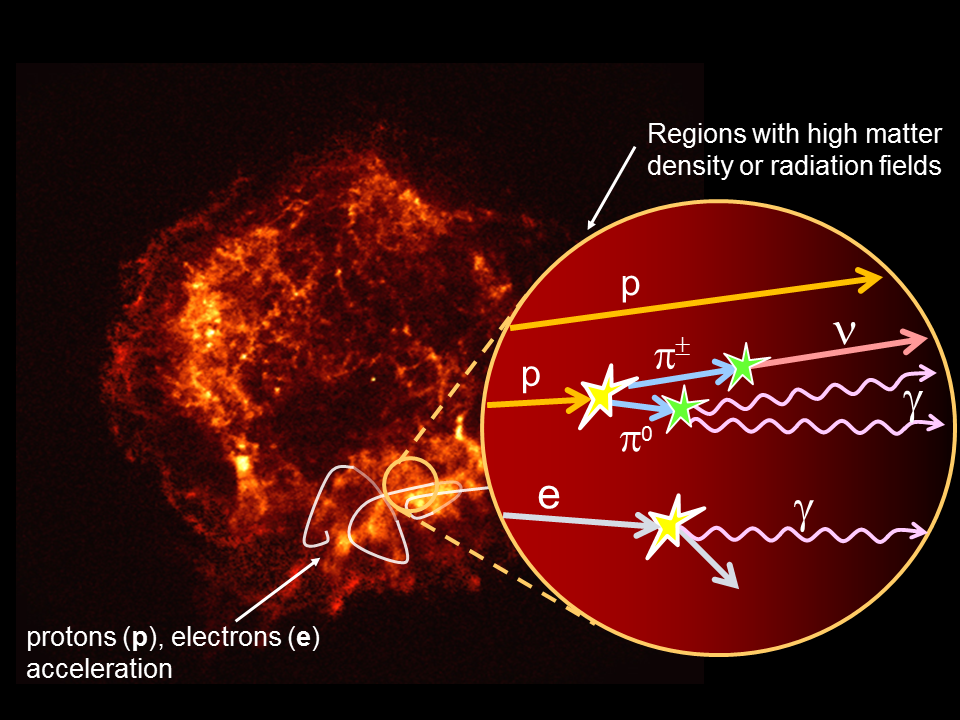}
\caption{\small Astrophysical sources can accelerate CRs (protons, electrons and nuclei). A~fraction of these particles diffuses outside the acceleration region, propagates in the galactic space (or intergalactic, if~the source is external to the Galaxy) and can reach the Earth. Electrically charged cosmic rays do not travel in a straight line due to magnetic field deflections. Another fraction of accelerated protons (or nuclei) can instead interact with matter ($pp$-mechanism) or radiation fields ($p\gamma$-mechanism) surrounding the source. In~this case, the~decay of a neutral secondary particle (mainly $\pi^0$) produces a $\gamma\gamma$-pair, while the decay of charged particles (mainly $\pi^\pm$) produces (anti)neutrinos. The~interaction of electrons with matter or radiation produces only $\gamma$-rays. Thus, the~detection of neutrinos from the direction of a source is a unique way to find out what are the accelerators of protons and~nuclei.}
\label{FigRC}
\end{center}
\end{figure}

The two major mechanisms for high-energy neutrino production, discussed just above, are usually referred to as $pp$-mechanism (Equation (\ref{eq:pp1})) and $p\gamma$-mechanism (Equation (\ref{eq:pg1})). They can apply to various situations in the cosmos. 
The interconnection between cosmic rays (CRs), $\gamma$-rays and neutrinos is sketched in Figure~\ref{FigRC}.

Let us examine now some differences between the $pp$- and the $p\gamma$-mechanism. The~main ones concern the~spectrum,
\begin{itemize}
\item[$(i)$] The $p\gamma$-mechanism is a process with a defined threshold.
For example, suppose having a photon target between the UV and X bands, say with energies of $\varepsilon \sim 0.1$ keV. For~the production of a $\Delta$-resonance, kinematics dictates:
\begin{equation}
E_p> \frac{m^2_{\Delta}}{4 \varepsilon_\gamma} =  \left(\frac{100\mbox{ eV}}{ \varepsilon_\gamma}\right)
\times 4\mbox{ PeV}
\label{eq:pg2},
\end{equation}
which, according to (\ref{eq:pp3}), corresponds to neutrinos  with minimal energy of $\sim 200$ TeV (see also Reference~\cite{a1}). In~this production mechanism, the~resulting neutrino (and very high-energy $\gamma$-ray) spectra will reflect the energy distribution of the target photons. 
\item[$(ii)$] The $pp$-mechanism is featured by a very important property of the hadronic interactions, namely the \revFirst{hypothesis of limiting fragmentation~\cite{yen}, to~which we refer in the following as \textbf{scaling}. A~detailed description of scaling variables, their definition, and~the application on cosmic ray physics can be found in Reference~\cite{ger}}.
 According to the scaling, the~secondary particle spectra corresponds quite strictly to the primary distribution. Consequently, if~the cosmic rays are power-law distributed, also the neutrinos and the very high-energy $\gamma$-rays will be power law distributed, with~(almost) the same slope. One commonly says that, in~this case, {\em the neutrino and gamma-ray spectra reflect the primary (cosmic ray) spectra}. 
In particular, if~some variant of Fermi acceleration mechanism applies, we would have
\begin{equation}
\Phi_\nu\propto E_\nu^{-\alpha} \mbox{ with }\alpha\sim 2.
\label{eq:pg3}
\end{equation}
\end{itemize}

Other differences between the $pp$-mechanism and $p\gamma$-mechanism concern the proportions of the neutrinos and gamma-rays.
Because a fraction of cosmic rays are nuclei, the~$pp$-collision from individual nucleons in the nuclei will still produce neutrinos, where $p\gamma$-collisions will mostly lead to photo-disintegration \revFirst{which  attenuates their energy~\cite{attenu}.} 
The latter process reduces the production of secondary neutrinos. 
Finally, in~the $p\gamma$-mechanism, a~relatively larger amount of neutral pions is produced than for the $pp$-mechanism, yielding more $\gamma$-rays than $\nu$'s.

\subsection{Connection of Neutrino and Gamma-Ray~Astronomies\label{sec:gn}}

The $pp$- and $p\gamma$- production of secondary particles described in Section~\ref{sec:hm} is usually referred as the {\em hadronic} mechanism for $\gamma$-rays and $\nu$ production.
However, a~direct connection between high-energy neutrinos and $\gamma$-rays requires that certain assumptions be fulfilled. Differences arises from the fact~that 
\begin{itemize}
	\item $\gamma$-rays can be produced also in the {\em leptonic} mechanisms in which only electrons are usually involved;
	\item $\gamma$-rays can be absorbed if the target is thick;
	\item when the $\gamma$-rays propagate for long distances, they are subject to absorption over background photons due to pair production
\begin{equation}\label{eq:ggee}
\gamma+\gamma_{\mbox{\tiny bkgr}} \to e^++e^-  \ .
\end{equation}
\end{itemize}

Concerning the last point, for~instance, the~opacity due to a thermal population with temperature $T_{\mbox{\tiny bkgr}}$ and distributed over a region with size $L$ is~\cite{celli}:
\begin{equation}\label{eq:ggee1}
\tau_{\mbox{\tiny pair}}=1.315 \times \frac{L}{\mbox{\small 10 kpc}} \times 
f\left( \frac{E_\star}{E_\gamma}\right) \mbox{ with } 
\left\{
\begin{array}{l}
E_\star=\displaystyle {m_e^2}/{T_{\mbox{\tiny bkgr}}}\\[1ex]
f(x)\approx - 3.68\ x\ \ln\!\left( 1 - e^{ - 1.52\ x^{0.89}} \right).
\end{array}
\right.
\end{equation}

Consider for example, the~ubiquitous CMB photons such that $E_\star=1.1$ PeV.
If the distance of propagation is 100 Mpc and the energy is
$E_\gamma=100$ TeV, we expect a survival fraction of $\exp(-\tau_{\mbox{\tiny pair}})=28$\%; but also for a Galactic distance $L=8.5$ kpc, a~photon of $E_\gamma=1$ PeV has  $\exp(-\tau_{\mbox{\tiny pair}})=39$\%. 
Therefore, the~attenuation effect is very important for propagation over extragalactic distances, but~it is not negligible for the highest energy galactic cosmic rays. 
When the contribution of  background infrared photons (due to \textit{reprocessed} star light) is included, the~Universe is expected to become opaque already at the TeV energy scale. 
Thus, at~the highest energies, the~connection between extragalactic gamma rays and neutrinos is indirect.
\revFirst{Detailed numerical codes for the description of more complex~situation where absorption occur 
are publicly available, see for example, those of {\tt nuFATE} and {\tt nuSQuIDS} codes in Reference~\cite{codici}.}

On the contrary, there are various cases when the connection is much more direct. For~instance, consider the case of young supernova remnants (SNRs), namely, the~mass of gas expelled by a supernova. This is considered as a plausible candidate for the acceleration of (galactic) cosmic rays.
Moreover, the~regions where core collapse supernovae explode are also sites of intense stellar formation activity, and~therefore, it often happens that SNRs are surrounded by  large amounts of target material, 
\revFirst{see References~\cite{targo1,targo2} and compare with Reference~\cite{targo3} that focusses only on particle physics aspects instead.}
This is the case of the SNR denoted as RX J1713.7-3946, that is just 1 kpc from the Earth and it is considered as a promising source of high energy neutrinos, whose neutrino flux can be predicted thanks to the observed gamma rays:
\revFirst{see Reference~\cite{jaime} for a first discussion, Reference~\cite{vils} for an accurate evaluation.}

In general, considering this type of galactic source, we expect~\cite{saak} a discernible signal in the neutrino telescopes of a 1 km$^3$-scale only if the $\gamma$-ray flux is larger than:
\begin{equation}
I_\gamma(>\mathrm{10\ TeV})=(1-2)\times 10^{-13}\,\mathrm{cm}^{-2}\,\mathrm{s}^{-1}.
\end{equation}
 
\begin{figure}[t]
\begin{center}
\includegraphics[width=0.8\textwidth]{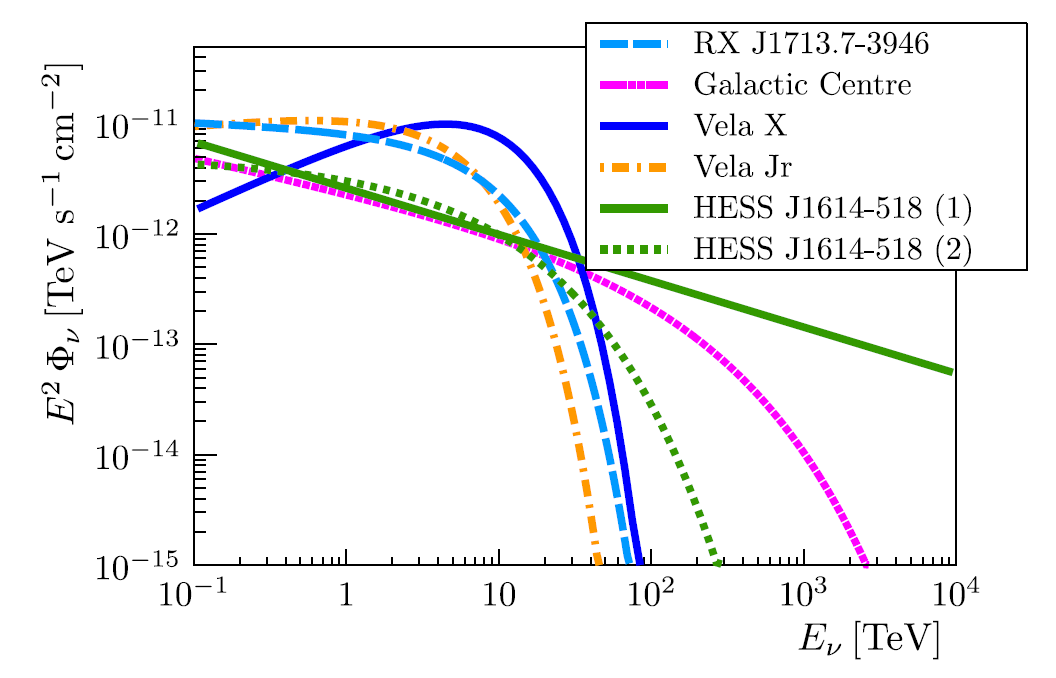}
\caption{\small  Prediction of the $\nu_\mu$ fluxes for different galactic sources used in the sensitivity studies of the KM3NeT collaboration~\cite{aiello}. }
\label{PSkm3}
\end{center}
\end{figure}

A detailed discussion of the signal of the neutrinos, derived following these principles     
from the $\gamma$-ray observations 
of certain specific galactic sources, can be found in Reference~\cite{aiello}, see Figure~\ref{PSkm3} from this reference. There, it is shown that for the KM3NeT detector (see Section~\ref{sec:km3}) an observation with 3$\sigma$ significance is possible in about six years of operation for sources as RX J1713.7-3946 and Vela~Jr.

\section{From Cosmic Neutrino Sources to the~Earth}

\subsection{Effects of Neutrino~Propagation\label{sec:osci}}
The relevance of neutrinos oscillations to the interpretation of cosmic neutrinos was 
remarked quite soon, see for example~\cite{gp69,bg77,bp78}. \revFirst{For instance,
in Reference~\cite{bg77} we read 
``if neutrino oscillations occur and cause the transition
$\bar\nu_{\mbox{\tiny e}} \rightleftarrows  \bar\nu_\mu$, 
then the flux of 
$\bar\nu_{\mbox{\tiny e}} $ increases. This effect  is particularly important for $p\gamma$ neutrinos''
while in Reference~\cite{bp78} we read
``if there exist more than two neutrino types with mixing of all neutrinos, cosmic
neutrino oscillations may result in the appearance of new type neutrinos, the~field
of which may be present in the weak interaction hamiltonian together with heavy
charged lepton fields.'' Actually, these observations still delimit the frontier of research in the field of cosmic neutrinos, for~the reasons we recall
here below and will elaborate in Sections~\ref{sec:dcore} and ~\ref{sec:glashow}.}

The vacuum oscillation phases that have been probed in terrestrial experiments can be parameterized as:
\begin{equation}\label{eq:osci1}
\varphi=\frac{\Delta m^2\ L}{2 E_\nu} = 3\times 10^4
 \, \frac{\Delta m^2}{7.4\times 10^{-5}\mbox{\,\small eV$^2$}}
 \, \frac{50\mbox{\,\mbox{\small TeV}}}{E_\nu} \,. 
 \frac{L}{\mbox{pc}}
\end{equation}

This means that \revFirst{vacuum} oscillations have to affect neutrino propagating over cosmic distances. 
{When we discuss astrophysical neutrinos, in~the energy range between hundreds of TeV and multi-PeV, produced by extragalactic sources (i.e., at~least at distances of Mpc), the~oscillation phase becomes order of $10^{10}$ or more. In~this case the only observable and meaningful physical quantity is the phase averaged oscillation. }
The minimal setup to analyze the effect on cosmic neutrinos is just the average value of the  vacuum probabilities when the oscillating phases are set to zero:
\begin{equation}\label{eq:osci2}
	P_{\ell\to\ell'}=\sum_{i=1}^3 | U_{\ell i}^2 | \ | U_{\ell' i}^2 | .
\end{equation}

This limit, when the oscillation probabilities reduce to constant values, is known as {\em Gribov-Pontecorvo} 
regime~\cite{gp69}.
\revFirst{Using the  most recent best fit values of the mixing angles~\cite{Esteban:2018azc}:
\begin{equation}\label{eq:osci3}
\{ \theta_{12},  \theta_{13},  \theta_{23},  \theta_{\mbox{\tiny CP}} \}=
\{  33.82^\circ, 8.61^\circ, 48.3^\circ, 222^\circ   \}
\end{equation}
 the resulting approximate values of the oscillation probabilities in Equation~(\ref{eq:osci2}) correspond to:
\begin{equation}\label{eq:osci4}
\mbox{matrix}(P_{\ell \to \ell'} ) =
\left(
\begin{array}{ccc}
0.548 & 0.185 & 0.267 \\
0.185 & 0.436 & 0.379 \\
0.267 & 0.379 & 0.354
\end{array}
\right) \mbox{ with }\ell,\ell'=\mbox{e},\mu,\tau
\end{equation}}

Given a flavor fraction of different neutrinos at the source, $f_\ell^0$, such that $0\le f_\ell^0\le 1$ and $ f_{\mbox{\tiny e}}^0+ f_\mu^0+f_\tau^0=1$,
one calculates the final flavor fraction at  Earth as:
\begin{equation}\label{eq:osci5}
f_\ell=\sum_{\ell=\mbox{\scriptsize e},\mu,\tau} P_{\ell \to \ell'} \ f_{\ell'}^0.
\end{equation}

(For a thorough investigation of the relevant parameters  and an assessment of the 
uncertainties, see References~\cite{natural,prompo}.)

The major consequence of Equation~(\ref{eq:osci5}) for cosmic neutrinos on Earth is that the spectra of the three   neutrino flavors are approximately the same, if~they have exactly the same power law distribution. 
This is particularly true in the case of muon and tau neutrinos, due to the parameters of oscillations, which display an approximate mu-tau exchange~symmetry.

Assuming in fact that neutrinos originate from $pp$ or $p\gamma$ production  (Section \ref{sec:hm}). In~astrophysical environments, the~pion decay chain is complete (i.e., also the muon decays).
\revFirst{According to Equation~(\ref{eq:pp2}), one has exactly a ($\nu_\mu +\bar \nu_\mu$) pair for each $\nu_e$ or $\bar\nu_e$, leading to flavor fractions at the source $f_{\mbox{\tiny e}}^0=1/3$,  $f_{\mu}^0=2/3$ and $f_{\tau}^0<10^{-5}$.
In all large apparata used to study cosmic neutrinos, there is no experimental method able to distinguish reactions induced by a neutrino or an anti-neutrino; thus, no separation between particles and anti-particles can be made. }

{
The flavor fractions can thus be written using a single parameter, $f=f_e^0$, as:
\begin{equation}\label{eq:osci6}
\{\, f_e^0, f_\mu^0,f_\tau^0 \ \}=\{ f , 1-f, 0 \} \ .
\end{equation}

With oscillation parameters given in Equation~(\ref{eq:osci4}), we obtain:
\begin{equation}\label{eq:osci7}
\{\, f_e, f_\mu,f_\tau \ \}=\{0.185 + 0.362 f, 0.436 - 0.251 f, 0.379 - 0.112 f \}
\end{equation}
that verify the properties mentioned just above. In~particular, for~$f=1/3$ the flavor fractions are 
$1/3+ \{-0.027, +0.019, +0.008 \},$ namely, they are very close to each other after oscillations.}

Tau neutrinos, in~this context of discussion, have a central role as they are not produced in charged mesons decay, either in atmospheric or astrophysical environments. 
The conventional component of atmospheric neutrinos from  pion and kaon decays, as~described in Section~\ref{sec:atnu}, is {\em mostly} composed by $\nu_\mu+\bar\nu_\mu$; at 
few 10-100 TeV one expects the onset of a new component due to charmed meson decays, that should instead be almost equally composed by electron and muon neutrinos (more discussion below).
Because at energy scales larger than 1 TeV atmospheric neutrinos are not significantly subject to oscillations in a path of length of the Earth diameter, we do not expect the presence of any atmospheric $\nu_\tau$.
Instead it is for sure that cosmic neutrinos are subject to neutrino oscillations, and~therefore tau neutrinos have to be present.
{The interpretation of tau neutrino events should be regarded with high confidence as due to high-energy neutrinos that have traveled upon cosmic distances.}  
See Reference~\cite{tau} for a quantitative and updated discussion of the~expectations.

{Note that, generally, neutrino and antineutrino events are not distinguishable and therefore 
are added together - the Glashow resonance events due to $\bar\nu_e$'s, discussed in Section~\ref{sec:glashow}, 
are an exception to this rule.}

\subsection{Neutrino Interactions in the~Earth\label{sec:abs}}  
The distance traveled by a neutrino in the Earth volume, before~reaching the detector, is:
\begin{equation}\label{eq:cs1}
\ell(\theta_{\mbox{\tiny N}}; d)=(R_\oplus-d) \cos\theta_{\mbox{\tiny N}}+ \sqrt{(R_\oplus)^2- (R_\oplus-d)^2\cdot (\sin \theta_{\mbox{\tiny N}} )^2},
\end{equation}
where $\theta_{\mbox{\tiny N}}$ is the nadir angle, $R_\oplus=6371$ km the average Earth radius.
{The quantity $d$ represents the depth of the detector (of the order of a few km at most)}.
 The formula is derived from basic trigonometric considerations. 


The Earth is not fully transparent to the highest energy neutrinos, due to the increase of the absorption   cross section of the matter, $\sigma_{\mbox{\tiny abs}}$, with~the increase of the neutrino energy $E_\nu$. 
The main contribution is due to the charged current (CC) interaction with nucleons 
$\sigma_{\mbox{\tiny abs}}\simeq \sigma_{\mbox{\tiny CC}}$
that can be modeled as deep inelastic scattering. The~neutral current interaction produces a neutrino with a reduced energy, and~this does not significantly affect the absorption cross section.  
{Detailed calculations (performed also through Monte Carlo methods) can account for the deformation of the arrival energy spectrum due to neutral current interactions and due to the $\nu_\tau$ regeneration effect~\cite{taureg}. The~propagation of the $\bar\nu_e$ flavor at 6.3 PeV is also affected by the resonant formation of the $W$ boson in $\bar \nu_e + e^-$ collisions, described in Section~\ref{sec:glashow}.}


The absorption coefficient of the neutrinos, according to the usual formula   $e^{-\tau}$, depends on the \textit{opacity factor} $\tau$, defined as:
\begin{equation}\label{eq:cs2}
	\tau\equiv \sigma_{\mbox{\tiny abs}}(E_\nu) \times z(\theta_{\mbox{\tiny N}}) \ .
\end{equation}

\begin{figure}[t]
\begin{center}
\includegraphics[width=0.49\textwidth]{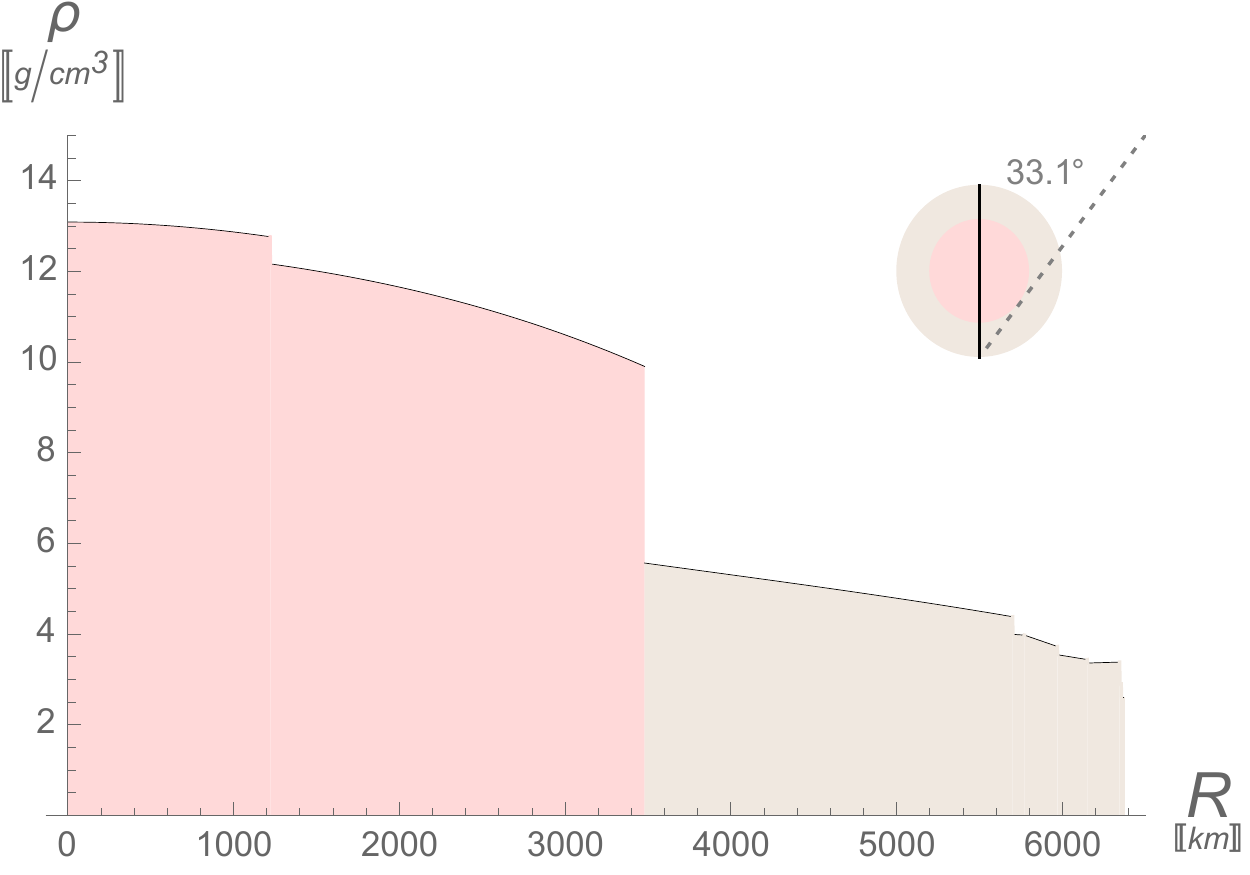}
\includegraphics[width=0.5\textwidth]{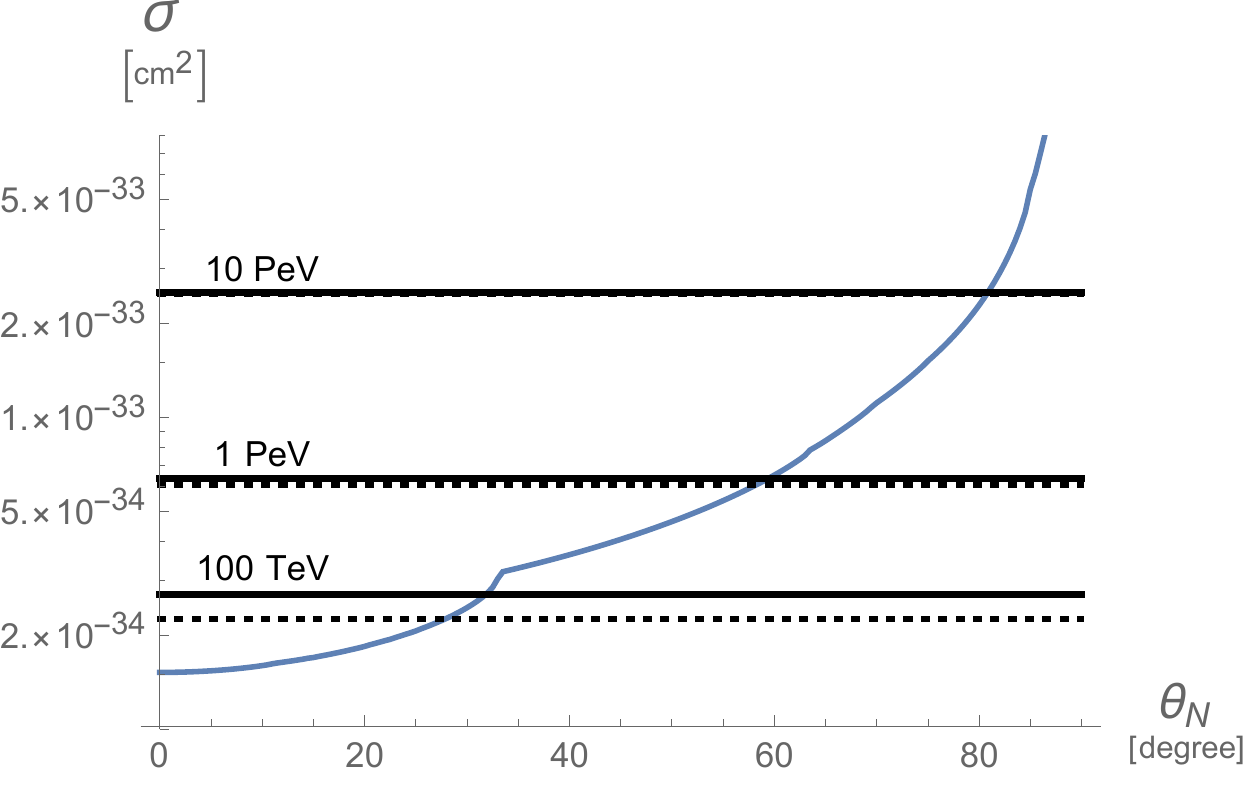}
\caption{\small Left panel: Earth's density in g/cm$^3$ according to the PREM model~\cite{PREM}; note the discontinuity between the mantle (brown area) and the core (red area) at a distance from the center of $R_{\mbox{\tiny core}}=3840$ km. The~limiting angle $\theta_{\mbox{\tiny N},\mbox{\tiny core}}=\arcsin(R_{\mbox{\tiny core}}/R_\oplus)=33.1^\circ$ is emphasized.
Right panel; three values of the neutrino (continuous lines) and antineutrino (dotted lines) cross sections, as~compared with the critical value of the cross section $\sigma_\star(\theta_{\mbox{\tiny N}})$ related to e-folding absorption, as~defined in the text.} 
\label{fig2}
\end{center}
\end{figure}

Here, the~\textit{column density} $z(\theta_{\mbox{\tiny N}})$, which corresponds to the number of nucleons per cm$^2$ that are crossed from a certain nadir angle, can be estimated by the formula
\begin{equation}\label{eq:cs3}
z(\theta_{\mbox{\tiny N}})=2 N_A \int_0^{R_\oplus\cos\theta_{\mbox{\tiny N}} }
\rho\left(\sqrt{x^2+ (R_\oplus\sin\theta_{\mbox{\tiny N}})^2 }\right) dx,
\end{equation}
where $N_A$ is the Avogadro number, $x$ is the coordinate along the neutrino path inside the Earth. Here, we neglected $d$ (and $h$) in the formula (\ref{eq:cs1}) for $\ell(\theta_{\mbox{\tiny N}};h,d)$, which is sufficient for our purposes.
The terrestrial density as a function of the distance from the center  $\rho(R)$, estimated using the PREM 
model~\cite{PREM} with an accuracy adequate to the current need, is shown in the left panel of Figure~\ref{fig2}. One can define the inverse of the column density as the 
{\em critical value of the cross section}
\begin{equation}\label{eq:cs4}
\sigma_\star(\theta_{\mbox{\tiny N}}) =1/z(\theta_{\mbox{\tiny N}}).
\end{equation}
In fact, when the absorption cross section 
satisfies the condition $\sigma_{\mbox{\tiny abs}}(E_\nu)=\sigma_\star(\theta_{\mbox{\tiny N}})$, the~opacity is $\tau=1$ and the absorption coefficients is $1/e\sim 0.37$.  
The right panel of Figure~\ref{fig2} compares $\sigma_\star(\theta_{\mbox{\tiny N}})$, depicted in blue and shown as a function of the nadir angle, with~the cross section given for three values of the energy, indicated in the black lines. 
The angles at which the absorption coefficients of the neutrinos is $1/e$ are,
$32^\circ$ for $E_\nu=100$ TeV,
$59^\circ$ for $E_\nu=1$ PeV,
$81^\circ$ for $E_\nu=10$ PeV. 
Therefore, at~very-high energies, we receive neutrinos arriving only from a limited patch of the sky, close to the local horizon.
Note that the rotation of the Earth changes the region of the sky seen in the course of the day, except~for 
the case of a detector located at the poles (such as IceCube) when this does not~change.

\section{High Energy Neutrino~Telescopes\label{sec:method}}

\subsection{Operating~Principles\label{sec:op}}
The basic structure of a high-energy neutrino telescope is a matrix of light detectors inside a transparent medium. This medium, such as ice or water at great~depths:
\begin{itemize} 
\item offers a large volume of target nucleons for neutrino interactions;
\item provides shielding against secondary particles produced by cosmic rays;
\item allows the propagation of Cherenkov photons emitted by relativistic {charged} particles produced by the neutrino interaction. 
\end{itemize}

\begin{figure}[t]
\begin{center}
\includegraphics[width=0.6\textwidth]{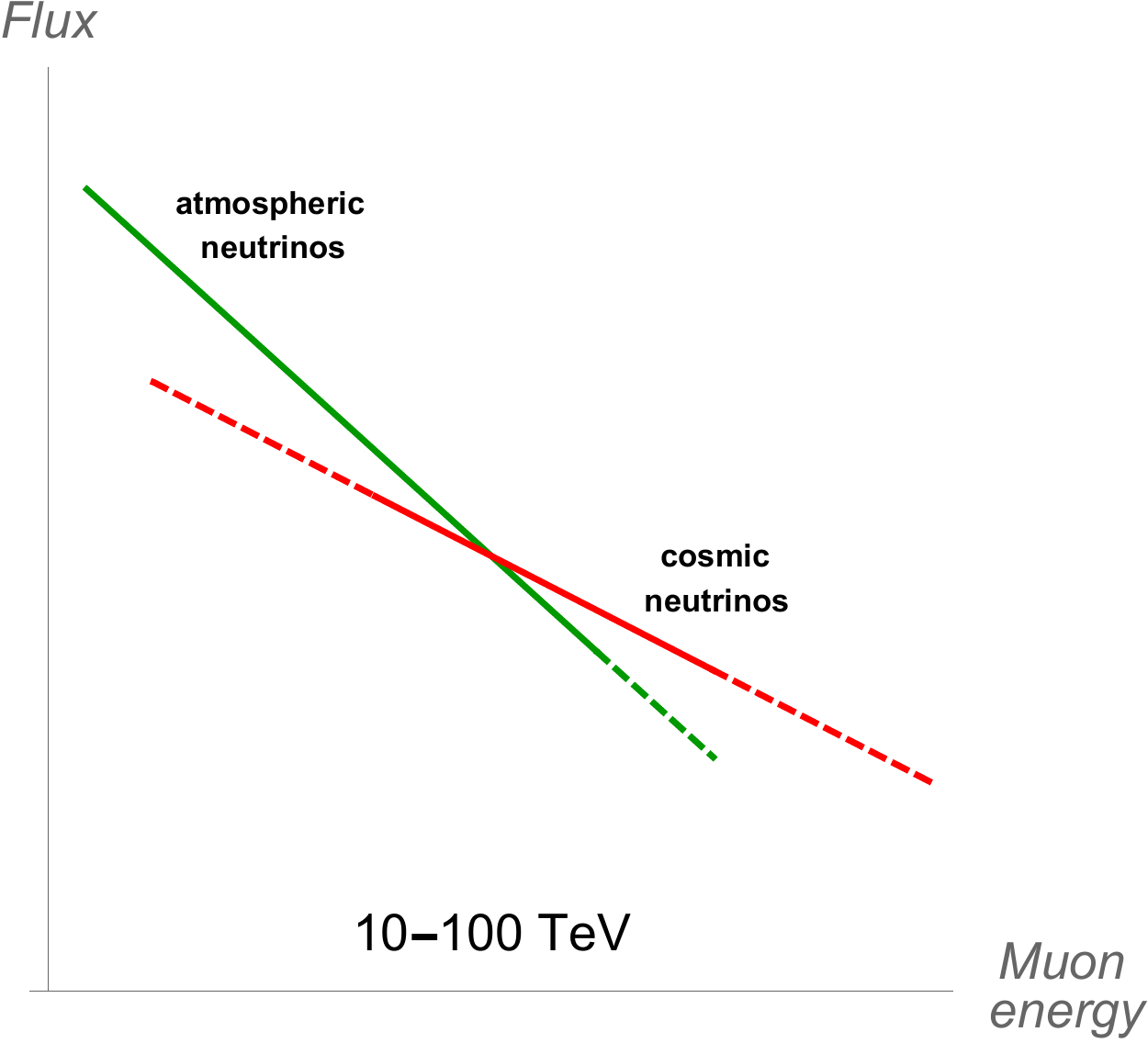}
\caption{\small Schematic representation of how a new ``hard''  muon neutrino component  of cosmic origin can stand out over the atmospheric neutrino at the highest energies, becoming observable. The~energy regions where the uncertainties 
are larger are represented with dashed lines. {For atmospheric neutrinos the uncertainty is due to the precise shape of the cosmic ray knee and the amount of prompt neutrinos; for cosmic neutrinos it is due to the production mechanism---moreover at low energies it is difficult to isolate them, while at high energies events are rare.} } 
\label{fig-n}
\end{center}
\end{figure}

The primary aim of these telescopes is the search for point sources and the {characterization of the neutrino flux produced by diffuse/unresolved sources}. While the idea in the former case is to stand out the atmospheric background in a specific direction of the sky, in~the latter case the idea is to rely fully on the study of the spectrum, in~the hope that a cosmic component, harder than the one due to atmospheric neutrinos, can eventually become visible. This is illustrated in Figure~\ref{fig-n} for the case of muon~neutrinos.  

Finally, a~neutrino candidate can be characterized in {\em time} and {\em direction} of arrival, and~it can be correlated with some temporal/spatial coincidence with an external triggers (such as that resulting from a $\gamma$-ray burst observation, from~a gravitational wave event, or~from other transients observed by space- or ground-based observations.) This last case is usually referred to as the \textit{multimessenger strategy}. {See Reference~\cite{murbar,Halzenmm} for recent reviews.}
The \textit{golden channel} for {point-like neutrino source searches} is represented by $\nu_\mu$'s, where a (sub) degree pointing can be attained in the energy of interest.

Neutrinos are also characterized by their {\em flavor}, {and each flavor and interaction mechanism induces different event topologies, as~discussed in Section~\ref{sec:topo}}. The~expected occurrence of neutrino oscillations give us a powerful manner to test the interpretation of the collected data (see Section~\ref{sec:osci}).

\subsection{Detectors All Around the~World\label{sec:GNN}}
The activities for the construction of a neutrino telescope started in the early 1970s with a joint Russian-American tentative.
Successively, the~efforts of the two communities turned toward an experiment in a Russian lake with an iced surface and in the ice of South Pole. At~the beginning of 1990's, European groups began the exploration of the Mediterranean Sea as a possible site for an underwater neutrino telescope.
{A detailed description of the experimental activities toward the realization of neutrino telescopes can be found in Reference~\cite{spie}}.

\subsubsection{IceCube\label{sec:IC}}
The IceCube experiment at the South Pole is, at~present, the~only running km$^3$-scale neutrino observatory.  
The instrumented detector volume is a cubic kilometer of highly transparent Antarctic ice instrumented with an array of 5160 Digital Optical Modules (DOMs).
{The IceCube array~\cite{icdet} is composed of 86 strings instrumented each with 60 Digital Optical Modules (DOMs). Among~them, 78 strings are arranged on a hexagonal grid with a spacing of 125 m, with~a vertical separation of 17 m between each DOM. The~eight remaining strings are deployed more compactly at the centre of the array, forming the DeepCore sub-detector~\cite{deepcore}. A~horizontal distance of 72 m separates the DeepCore strings with a vertical spacing of 7 m between each PMT.}

The DOMs are spherical, pressure-resistant glass housings each containing a 25\,cm diameter photomultiplier tube (PMT) plus electronics for waveform digitization, and~vertically spaced 17\,m from each other along each string. High quantum efficiency PMTs are used in a denser sub-array located in the center of the detector. This sub-array, called DeepCore, enhances the sensitivity to low energy neutrinos. Data acquisition with the complete configuration started in May 2011~\cite{icsyst}.

\subsubsection{ANTARES\label{sec:antares}}
In the sea, to~minimize the noises induced by external agents, a~telescope must be located far enough from continental shelf breaks and river estuaries. At~the same time, the~detector should be close to scientific and logistic infrastructures on shore. With~such requirements, the~Mediterranean Sea offers optimal conditions on a worldwide~scale.

The {ANTARES} detector~\cite{antadet} was completed in 2008, after~several years of site exploration and detector R\&D. The~detector is located at a depth of 2475 m in the Mediterranean Sea, 40\,km from the French town of Toulon. It comprises a three-dimensional array of 885 optical modules (OMs) looking 45$^\circ$ downward and distributed along 12 vertical detection lines. An~OM consists of a 10$^{\prime\prime}$ PMT housed in a pressure-resistant glass sphere together with its electronics~\cite{antadom}. The~total length of each line is 450 m; these are kept taut by a buoy located at the top of the line. The~instrumented volume corresponds to about 1/40 of that of IceCube, but~due to the water properties, {and the denser configuration of OMs, its} sensitivity is relatively larger at energies below 100 TeV. In~addition the Galactic center is below the horizon $\sim$70\,\% of the~time.

\subsubsection{KM3NeT\label{sec:km3}}
KM3NeT is a research infrastructure that will house the next generation neutrino telescope in the Mediterranean Sea~\cite{km3loi}. 
KM3NeT will consists of two different~structures. 

{The KM3NeT/ARCA telescope is in construction about 100 km off-shore Portopalo di Capo Passero (Sicily), at~a depth of 3500 m. The~DOM spacing along the DU is 36 m with an average distance among DUs of 90 m. ARCA will consists of two building blocks of 115 vertical detection units (DUs) anchored at a depth of about 3500 m. The~telescope will have an instrumented volume slightly larger than that of IceCube.
The main scientific objectives are the study of astrophysical potential neutrino sources, with~particular attention for galactic ones~\cite{aiello}. The~golden channel is the detection of long track muon produced in charged current $\nu_\mu$ interactions.
For this kind of events an angular resolution better than 0.1$^\circ$ and an energy resolution of 30\% in log(Energy) are reached for neutrino energies greater than 10 TeV~\cite{km3loi,km3intr}.

The KM3NeT/ORCA detector is located on the French site  off-shore Toulon and at a depth of 2450 m. The~main scientific objectives are the determination of the neutrino mass hierarchy and the searches for dark matter. The~two KM3NeT facilities will also house instrumentation for Earth and Sea sciences for long-term and on-line monitoring of the deep-sea environment.
The DOM spacing along the DU is 9 m and the average distance among the DUs is 23 m. It will consists of one building block, with~an instrumented mass of 8 Mton.}

Both KM3NeT structures will use the same DUs equipped with 18 optical modules, with~each optical module comprising 31 small PMTs. The~technical implementation and solutions of ARCA and ORCA are almost identical, apart from the different spacing between~DOMs.

\subsubsection{GVD\label{sec:GVD}}
{Finally, many years of experience with smaller neutrino telescope installations in Lake Baikal in Siberia, have led to the Baikal- Gigaton Volume Detector (GVD) project, aiming at a detector instrumenting a cubic-kilometer of water~\cite{gvd}. 
Baikal-GVD is formed by a three-dimensional lattice of optical modules, which consist of PMTs housed in transparent pressure
spheres. They are arranged at vertical load-carrying cables to form strings. The~telescope has a modular structure and consists of functionally independent clusters. Each cluster is a sub-array that comprises 8 strings. Each cluster is connected to the shore station by an individual electro-optical cable. The~first cluster with reduced size was deployed and operated in 2015. In~April 2016, this array has been upgraded to the baseline configuration of a GVD-cluster, which
comprises 288 optical modules attached along 8 strings at depths from 750 m to 1275 m. In~2017--2019, four additional GVD-clusters were commissioned, increasing the total number of optical modules up to 1440 OMs. As~part of phase 1 of the Baikal-GVD construction, an~array
of nine clusters will be deployed until 2021. GVD will primarily target neutrinos in the multi-TeV range. First results have recently been reported~\cite{GVDicrc}. }

\section{Topologies of the~Events\label{sec:topo}}

{In neutrino telescopes, one can distinguish between two {\emph{main}} event classes---events with a long \textbf{track} due to a {through-going} muon (induced by $\nu_\mu$ charged current (CC) interactions), and~events with a \textbf{shower}, without~the presence of a muon.
{The special case of $\nu_\mu$ CC interactions inside the detector will contain a shower originating in the interaction vertex
and an accompanying track as well.}

A high-energy electron resulting from a CC $\nu_e$ interaction radiates a photon via bremsstrahlung after a few tens of cm of water/ice (the radiation length in water is $\sim$ 36 cm): this process leads to the development of an electromagnetic cascade (the $shower$).
Showers are induced as well by neutral current interactions and by $\nu_\tau$ CC interactions occurring inside the instrumented volume of the detector. Tau neutrino interactions can produce also a particular  event topology, with~events called in this review \textit{double core}.
} 

Neutrino and anti-neutrino reactions are not distinguishable; thus, no separation between particles and anti-particles can be made.
A shower of particles is produced {in proximity of the  interaction vertex in all neutrino interactions}. 
However, for~CC $\nu_\mu$, often only the muon track is detected, as~the path length of a muon in water exceeds that of a shower by more than 3 orders of magnitude for energies above 2\,TeV.
Therefore, such an event might very well be detected even if the interaction has taken place several km outside the instrumented volume, in~the Earth's crust or in the surrounding transparent medium, provided that the muon traverses the detector. 
More in detail, the~muon range $R$ can be described in the continuous-energy-loss approximation as:
\begin{equation}\label{eq:mu1}
\frac{dR}{dE_\mu}=-\frac{1}{\alpha+\beta E_\mu}
\mbox{ with }\left\{
\begin{array}{l}
\alpha\approx 2\times 10^{-6}\ \frac{\mbox{TeV\ cm}^2}{\mbox{g}},\\[2ex]
\beta\approx 4\times 10^{-6}\ \frac{\mbox{cm}^2}{\mbox{g}}.
\end{array}
\right.
\end{equation}

The $\alpha$ coefficients describes the well-known fact that low energy muons {\em (minimum ionizing particles}, mip)
loose about 2 MeV/cm in water, while the second coefficients
$\beta$ describes that the energy loss processes become `catastrophic' at high energies. This implies the travel distance in water or ice ($\rho\simeq 1$ g cm$^{-3}$):
\begin{equation}\label{eq:mu2}
d(E_\mu,E_{\mbox{\tiny thr.}})\approx  D \times  \log\left(\frac{1+{E_\mu}/{\varepsilon}}{1+E_{\mbox{\tiny thr.}}/{\varepsilon}}\right)
\mbox{ with }\left\{
\begin{array}{l}
\displaystyle
D = \frac{1}{\rho\cdot \beta} \approx 2 \mbox{ km}\\[3ex]
\displaystyle
\varepsilon=\frac{\alpha}{\beta}\approx \mbox{\small 0.5 TeV}.
\end{array}
\right.
\end{equation}

\begin{figure}[t]
\begin{center}
\includegraphics[width=0.8\textwidth]{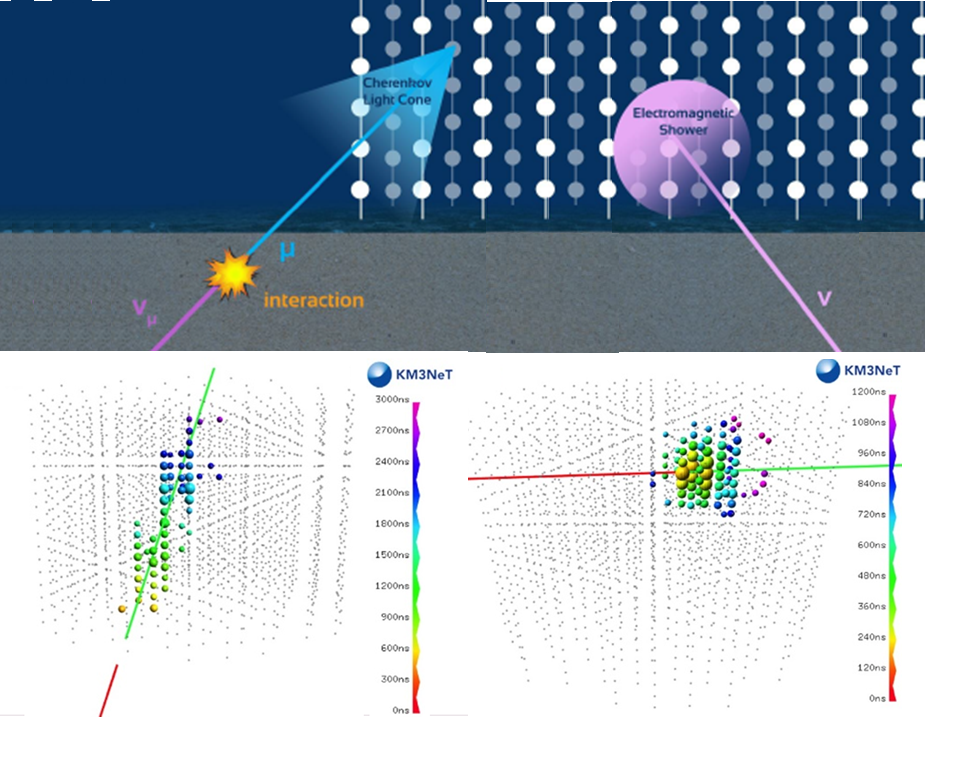}
\end{center}
\caption{\label{fig:trsh} 
\small 
 Neutrino interactions give two different types of events: $\nu_\mu$ charged current (CC) interaction that produces a muon (left) and $\nu_e$ CC interaction that produces an electron inducing an electromagnetic cascade (shower). Neutral current (NC) interactions also produce showers. In~this sketch, provided by the KM3NeT Collaboration, each point represents an optical module (OM), spaced about 100 m from each other. In~gray are the OMs not interested by {optical} photons (mainly from the Cherenkov effect) produced by the event. The~colors show the temporal sequence of the hit OMs---the red color indicates the beginning of the event. }
\end{figure}

The properties of showers (total visible energy, a~rough estimate of the neutrino direction) are obtained if the interaction occurs inside {(or very close to)} the instrumented volume.
Figure~\ref{fig:trsh} shows a sketch of neutrino interactions giving the two different types of events. 
\unskip

\subsection{Passing-Events\label{sec:passing}}
{\em Passing-events} are due to tracks induced by $\nu_\mu$ and $\bar\nu_\mu$  CC interactions around the detector and passing through the detector. The~same signal is called (depending on the experiment) also through-going muons, up-going muons, passing muons, tracks, and~so forth emphasizing a feature or another one.
{(As recalled above, there is also the possibility of  track events that start inside the detector.)}

The angular precision for passing-events depends on the muon energy and detection medium. 
Deep ice is more transparent than seawater. Thus, the~same instrumented volume of ice corresponds to a larger effective volume than in seawater. On~the other hand, the~effective scattering length for ice is smaller than water. This is a cause of a larger degradation of the angular resolution of detected neutrino-induced muons in ice with respect to water. Passing-events are better reconstructed in seawater  than in ice.
{Refer to Reference~\cite{a5} for a detailed discussion of differences between water and ice properties.}
The typical angular resolution of tracks in ice, $\delta\theta_{\mbox{\tiny passing}}\sim 0.5-1^\circ$, while it is foreseen to reach $0.1^\circ$ in the KMeNeT~detector.

Passing-events are mostly selected from the direction of the Sky below the detector, in~order to keep under control the background of atmospheric muons. Therefore, the~neutrinos that have to cross a large fraction of the Earth, above~few 100 of TeV, are subject to absorption (Section \ref{sec:abs}); the optimal region of observation is therefore a crown of azimuthal directions, just below the horizon of the~detector.

\subsection{Contained~Events\label{sec:hese}}
The first operating km$^3$-class telescope, IceCube, has demonstrated that also events partially or fully contained in the detector can be used to observe high-energy cosmic~neutrinos. 

The method rely on the request that the interaction point, and~possibly the whole event, is located into the instrumented volume. 
In this manner, the~background can be suppressed, and~this is even more effective for electron type events, that are less abundant among atmospheric neutrinos. 
The ratio of the through-going events with respect contained ones is roughly given by
\begin{equation}\label{eq:mu3}
\frac{N_{\mbox{\tiny throug.}}}{N_{\mbox{\tiny contained}}}\sim \frac{\mbox{area} \times D}{\mbox{volume}}
\end{equation}
where a detector with typical dimension $L$ has an area $\sim L^2$ for muon detection and a volume $L^3$ for contained event detection. 
The scale distance D is defined in Equation~(\ref{eq:mu2}).
The ratio is just $\sim D/L$ and this is not small as long as the detector has physical dimension $L\sim D$ namely, kilometer size, just what has been attained by~IceCube.

This method allows the telescope to observe also events coming from {\em above} the detector, differently from the previous method.
However, the~risk of background contamination from atmospheric muons does exist, especially at the lowest energies. 
For zenith angles $\theta < 70^\circ$ and $E_\nu >$ 100 TeV the atmospheric muon self-veto (Section \ref{sec:atmu}) can help to suppress the atmospheric neutrino~background.

\subsection{Double Core~Events\label{sec:dcore}}
{The presence of $\nu_\tau$ would be an unequivocal signature of a cosmic neutrino. The~direct production  in cosmic ray collision
is very suppressed. High-energy atmospheric  neutrinos do not have 
enough time to undergo flavor transformation. Thus, 
with  high confidence,  $\nu_\tau$ are  produced through three flavor oscillations over cosmic scale distances (Section \ref{sec:osci}). }

The identification of a $\nu_\tau$ relies on the extrapolation on macroscopic scales of the principle used by OPERA, namely, the~displacement between the point where the tau lepton is produced (by a CC $\nu_\tau$ interaction) and the point where the charged tau decays.
There are two different implementations of this principle. In~the first one, different regions of a neutrino telescope (namely, different strings or set of phototubes) see the two interaction vertexes. 
In the second one, using the fast time response of each individual phototube, the~experiment is able to distinguish light emission occurred in two successive processes  (the $\nu_\tau$ interaction, the~$\tau$ decay).
These two principles are known as {\em double bang} \cite{2b} and {\em double pulse} \cite{2p} respectively---and collectively are called {\em double core events} {in the present review paper}.

\subsection{Glashow~Resonance\label{sec:glashow}}
An exciting possibility is the observation of events caused by $\bar \nu_e$  interacting with atomic electrons, and~producing an on-shell $W$ boson via $\bar\nu_e+ e\to W$. The~cases when the $W$ decay in a pair of quarks yields the whole energy of the initial neutrino, amounting to
\begin{equation}\label{eq:mu4}
E_{\mbox{\tiny Glashow}}=\frac{m_W^2}{2 m_e}=6.3\mbox{ PeV}
\end{equation}

{This process yields a conventional observable shower; however, due to the large cross-section, this would incredibly enhance the opportunity to observe PeV-energy $\bar\nu_e$. An~events at a such energy, yet unseen, is called {\em Glashow resonance} \cite{glashow}.}

Seeing events due to Glashow resonance would be an excellent way to probe of the extension of the cosmic neutrinos spectrum at high energies, and~in the long run, also a test of the type of neutrino source (whether $pp$ or $p\gamma$). See Reference~\cite{palla} for an updated, quantitative~discussion.

\subsection{Effective~Areas}
For each event topology, the~calculation of the expected number of neutrino signals $N_i$, given a flux of neutrinos of type $\ell$, is usually performed by means of {\em effective areas} $A_{\ell \to i}$. 
If the flux is constant in time and the observation time is $T$, it is possible to write symbolically:
\begin{equation}\label{eq:effa}
 N_i={4\pi} T \int dE_\nu \ A_{\ell \to i}(E_\nu)\  {\Phi_\ell} 
\end{equation}

The effective area incorporates the neutrino cross section, the~number of target particles in the detector, the~angular and the energy response, the~detector inefficiencies, the~cuts implemented in the analysis. Usually, it is a growing function of the energy of the incoming neutrino. In~the IceCube detector the effective areas for the main classes of events, the~passing (muon) events and the contained events, are about some tens of m$^2$ in the relevant energies. The~effective area for contained events at high-energies increases mostly because of the cross section growth with neutrino energy, $\sigma \sim E_\nu^{0.3-0.4}$; see Reference~\cite{gandhi} for a useful and accurate~discussion.

\section{Background~Processes}
The cosmic neutrino sky as seen on Earth is not background free, due to the presence of the atmospheric neutrinos and secondary atmospheric~muons.

\subsection{Atmospheric~Neutrinos\label{sec:atnu}}

In the region of energy where atmospheric neutrino oscillations have been discovered, $E_\nu \sim$ GeV, the~spectrum of the neutrinos is distributed as $E_\nu^{-\alpha}$ with $\alpha\sim 2.7$.
In fact, due to the mentioned property of {\em scaling} of the hadron-hadron interactions, atmospheric neutrinos have (approximatively) the same power-law index of the primary cosmic rays at the corresponding energies.
{Neutrino telescopes have been able to measure the flux and energy spectrum of atmospheric $\nu_\mu$ \cite{ICAmu,ANTAmu} and atmospheric $\nu_e$ \cite{ICAe}.}

This behavior changes at the higher energies of interest for the search of cosmic neutrinos, because~most secondary muons in Equation~(\ref{eq:pp2}), will touch ground before decaying. This reduces the number of atmospheric $\nu_\mu$ and depletes electron neutrinos: indeed, the~electron/muon neutrino fraction decreases from 1/2 to 1/30, because~the muon is not allowed to decay freely. 
Pions of similar energy travel 100 times less before decaying; however, they have a nuclear interaction cross section of the order of 
$\sigma_\pi\sim \pi\times \mbox{fm}^2$ with nucleons, which implies an interaction length $m_n/( \sigma_\pi \times \rho_{\mbox{\tiny air}})$ of comparable size for a density of $\rho_{\mbox{\tiny air}}\sim 10^{-3}$ g/cm$^3$. For~this reason, spectrum of atmospheric neutrinos from pion interactions becomes steeper,  $E_\nu^{-\alpha}$ with $\alpha\sim 3.7$, since only a fraction of pions is free to decay before interacting.
This is the behavior of the atmospheric component shown in Figure~\ref{fig-n}.

\begin{figure}[t]
 \includegraphics[width=\textwidth,keepaspectratio]{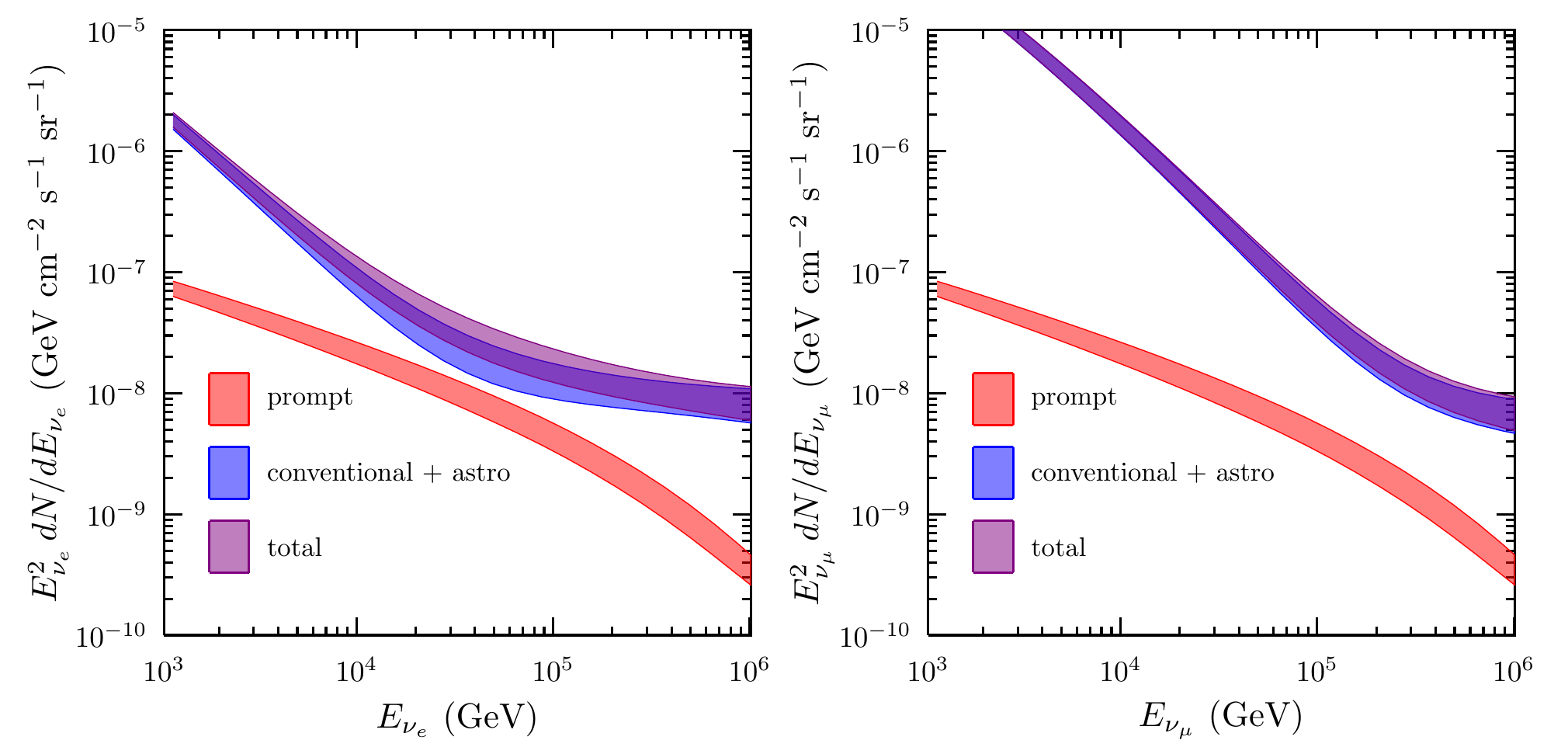} 
 \caption{The expected electron (left panel)  
and muon neutrino (right) flux components in the energy range from 1 TeV to 1 PeV.
 Prompt neutrinos are shown separately and summed with all other components.
From Reference~\cite{prompo}.
}
 \label{fig:convastro}
\end{figure}

It is important to recall that at some 10--100 TeV one expects the onset of a new component (to date unobserved) due to the production of charmed mesons, that, having a very short lifetime, decay immediately after production leading to a spectrum as $E_\nu^{-\alpha}$ with $\alpha\sim 2.7$ and with an approximately equal amount of electron and muon neutrinos. 
A very interesting possibility to emphasize and possibly observe the prompt contribution 
is to identify a clean sample of {\emph{non-muonic neutrinos}} in the region of 10-few 10 of TeV, as~discussed in Reference~\cite{prompo}.
{In this paper one finds a description of the model of atmospheric neutrinos, of~the one of cosmic neutrinos
(based on the hypothesis of pp-collisions and upgraded thanks to IceCube through-going muon data), and~an assessment of 
their uncertainties. For~an illustration of the principle, see Figure~\ref{fig:convastro}.}

\subsection{Atmospheric~Muons\label{sec:atmu}}

\begin{figure}[t]
\begin{center}
\includegraphics[width=0.6\textwidth]{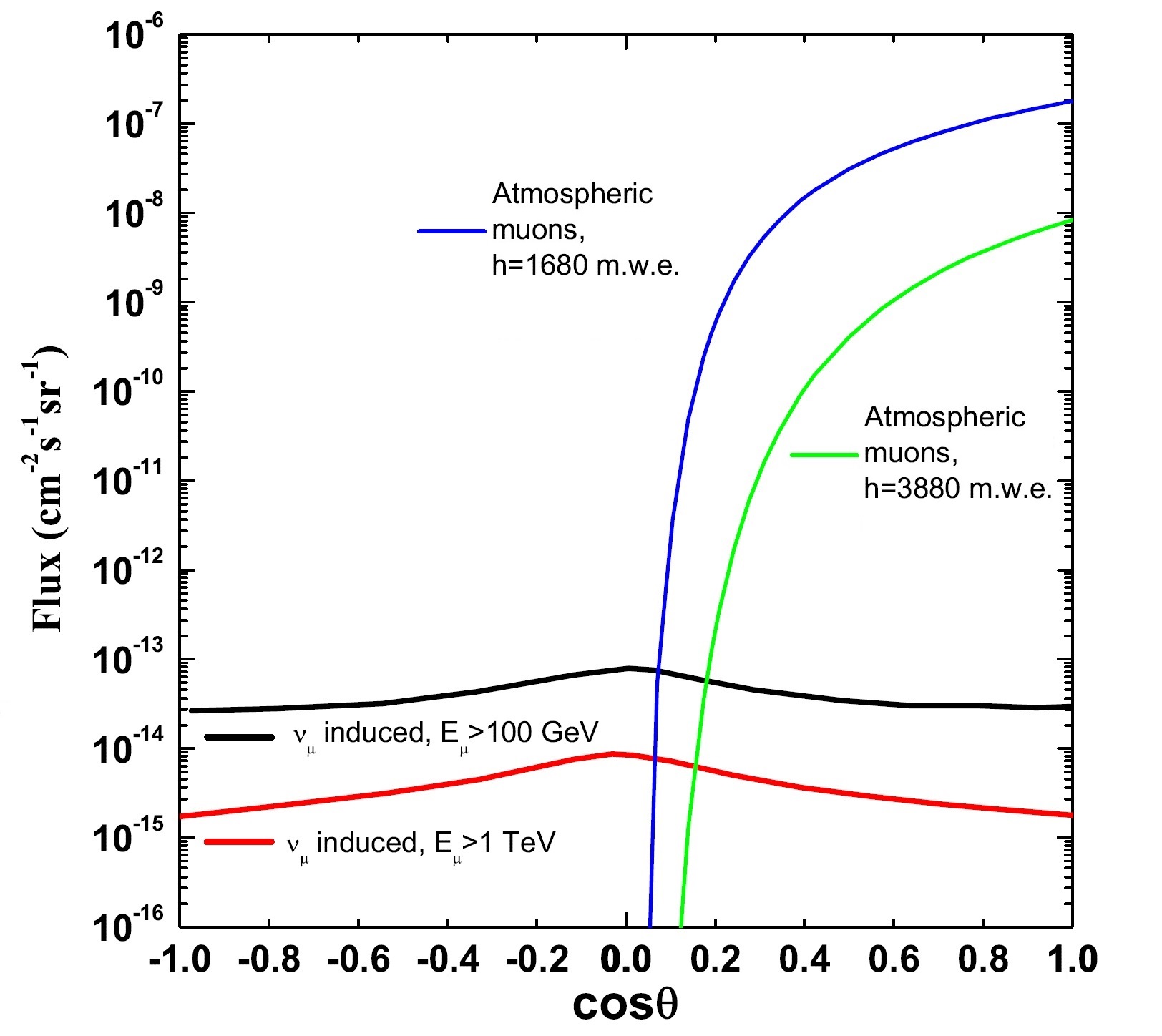}
\end{center}
\caption{\label{fig:atmunu} 
\small Flux as a function of the 
 cosine of the \textit{zenith angle} $\theta$ of: (i) atmospheric muons for two different depths; (ii) muons induced by CC interactions of atmospheric $\nu_\mu$, for~two different muon energy thresholds $E_\mu$. Upgoing (down-going) events have $\cos\theta<0 \ (>$$0)$ \cite{mupara}.}
\end{figure}

Atmospheric muons can penetrate the atmosphere and up~to several kilometers of ice/water, and~they represent the bulk of reconstructed {background events in any large volume neutrino detector~\cite{ICmuon,NEMOmuon,ANTAmuon}.} Neutrino detectors must be located deep under a large amount of shielding in order to reduce the background.
The flux of down-going atmospheric muons exceeds the flux induced by atmospheric neutrino interactions by many orders of magnitude, decreasing with increasing detector depth, as~shown in Figure\,\ref{fig:atmunu}.

Atmospheric muons can be used for a real-time monitoring of the detector status and for detector calibration~\cite{ANTAcali}. However, they represent a major background source: downward-going particles wrongly reconstructed as upward-going and simultaneous muons produced by different cosmic ray primaries could mimic high-energy neutrino interactions, see~\cite{mupara} for a~discussion.

In general, atmospheric neutrinos are indistinguishable from astrophysical neutrinos. The~presence of a muon can help to classify a contained interaction as due to atmospheric $\nu_\mu$ instead of a cosmic $\nu_\mu$. This occurs when neutrino energy is sufficiently high and the zenith angle sufficiently small that the muon produced in the same decay as the neutrino {has a high probability} to reach the detector~\cite{gais,gaisnew}. 
In this case, the~atmospheric neutrino provides its own self-veto. 
In general, the~atmospheric neutrino passing rate can be evaluated with simulations by including other high-energy muons produced in the same cosmic-ray shower as the neutrino. In~this way, the~method can be extended to $\nu_e$'s~\cite{gaisnue}. In~practice, the~passing rate is significantly reduced for zenith angles $\theta < 70^\circ$ and $E_\nu >$ 100 TeV.
This method was used first by the IceCube collaboration to select the high energy starting events (Section \ref{sec:hese}).

\section{{The Observational Status of High-Energy Neutrinos~Astronomy}}\label{sez:EHnu}

 The IceCube experiment has observed neutrinos of astrophysical origin in different ways~\mbox{\cite{halzsurv,AhlersH}}---first with the \textit{High Energy Starting Event (HESE)} sample; then with passing events; finally with a coincidence with electromagnetic~signals.

\subsection{Contained~Events}\label{sez:hese2}
The first detailed observation of an excess of high-energy astrophysical neutrinos over the expected background has been reported by IceCube using data collected from {May 2010 to May 2012 and with 662\,days} live time~\cite{10Ic13}.  
This sample is continuously updated, and~as of this writing (after ICRC 2019), results up to early 2018 are available for a total live time of 2635 days of live time. 
The high-energy neutrino candidates have been selected with the requirement that the interaction vertex is contained within the instrumented ice volume, without~any signal on the PMTs located on the top or sides of the detector. 
In such a way, the~edges of IceCube are used as a veto for down-going atmospheric muons and atmospheric neutrinos, Section~\ref{sec:atmu}.

The deposited energy $E_{dep}$ in the detector is derived from the number of photoelectrons (p.e.) in the optical modules and, in~turns, the~true energy $E_\nu$ of the neutrino is estimated with the help of Monte Carlo simulation techniques. An~event with 6000 p.e. corresponds to a deposited energy of ${\sim}30$\,TeV. 

\begin{figure}[t]
\begin{center}
\includegraphics[width=1.0\textwidth]{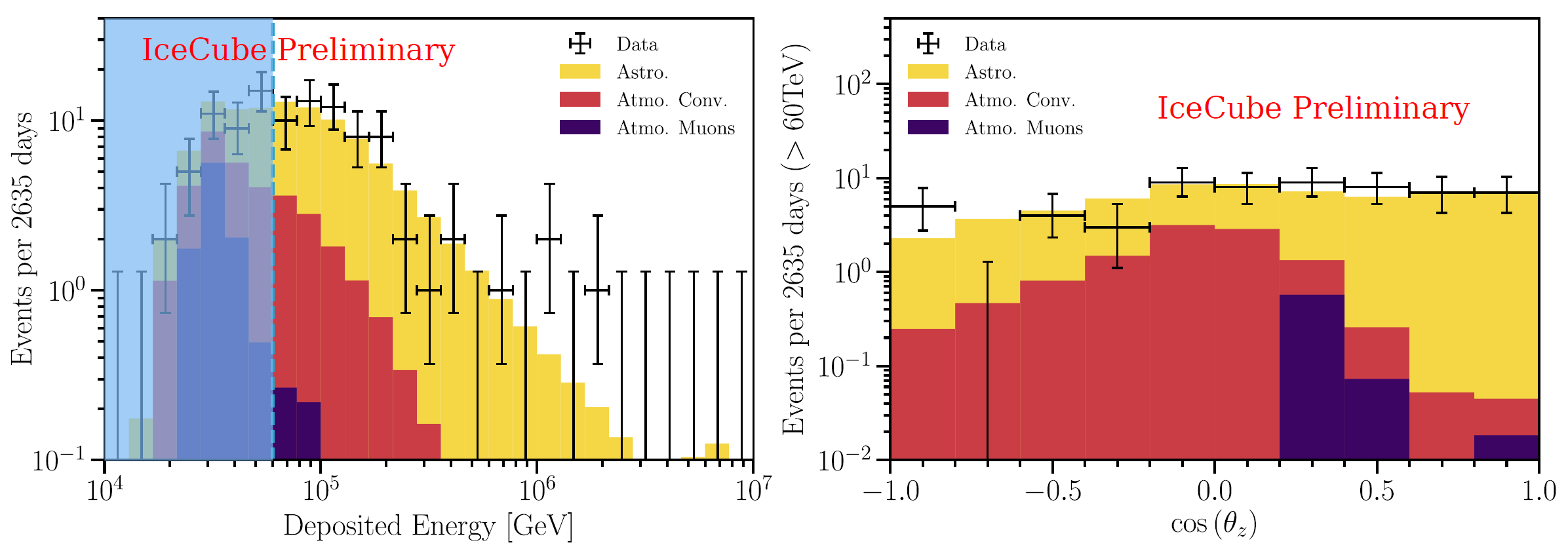}
\end{center}
\caption{\label{fig:10ICene}
\small 
Deposited energies 
 $E_{dep}$ (left panel) and arrival directions (right panel) of observed IceCube HESE ({crosses}), compared with predictions. The~sample refers to 7.5 years of data. 
The best-fit expectation is shown as a stacked histogram with each color specifying a flux component---astrophysical (golden), conventional atmospheric (red), and~penetrating muons (purple); the best fit prompt normalization is zero and is not shown.
Events below 60 TeV (light blue vertical line) are not included in the fit, but~one sees good data-MC agreement extending into this energy range. 
From the arXiv version of~\cite{IC1004}.} 
\end{figure}

Figure~\ref{fig:10ICene} shows the distribution of the deposited energy (left plot) and of the cos(declination) (right plot) for HESE in the 7.5-year sample. 
The additional contribution in the data sample with respect to the background corresponds to a diffuse astrophysical signal, with~topologies compatible with neutrino flavor ratio $\nu_e:\nu_\mu:\nu_\tau {\sim}1:1:1$, as~expected for a cosmic signal~\cite{ICflaratio}. 
The best fit to the data above 60 TeV~\cite{IC1004} yields, for~all neutrino flavors ($6\nu$ means the sum over neutrino and antineutrinos of all flavors),
\begin{equation}\label{eq:10hese}
\frac{d \Phi_{6\nu}}{dE}= (6.45^{+1.46}_{-0.46})\cdot 10^{-18} 
\biggl( \frac{E}{100 \textrm{ TeV}} \biggr)^{-(2.89\pm{0.2})} \textrm{ GeV}^{-1}\textrm{ cm}^{-2}
\textrm{ s}^{-1}\textrm{ sr}^{-1}  \ .
\end{equation}

Most of the signal originates primarily from the Southern hemisphere, where neutrinos with $E_\nu\gg 100$\,TeV are not absorbed 
by Earth. The~poor angular resolution (${\sim}15^\circ$) of showering events prevents the possibility of accurate localization in the sky of the parent neutrino's direction. 
To identify any bright neutrino sources in the data, the~usual maximum-likelihood clustering search has been used, as~well as searches for directional correlations with TeV $\gamma$-ray sources.
No hypothesis test has yielded statistically significant evidence of clustering or~correlations.

\subsection{The Passing~Muons}\label{sez:passing}
A second IceCube sample that evidences a diffuse presence of cosmic neutrinos corresponds to CC upgoing muon neutrino events~\cite{ic2y,6yr}. 
The field of view for these events is restricted to the Northern hemisphere. This analysis has recently been extended with data collected up to December 2018, equivalent to 10 y of live time~\cite{IC1017}. 
This last sample contains $\sim$650,000 muon neutrino~candidates.

\begin{figure}[t]
\begin{center}
\includegraphics[width=0.65\textwidth]{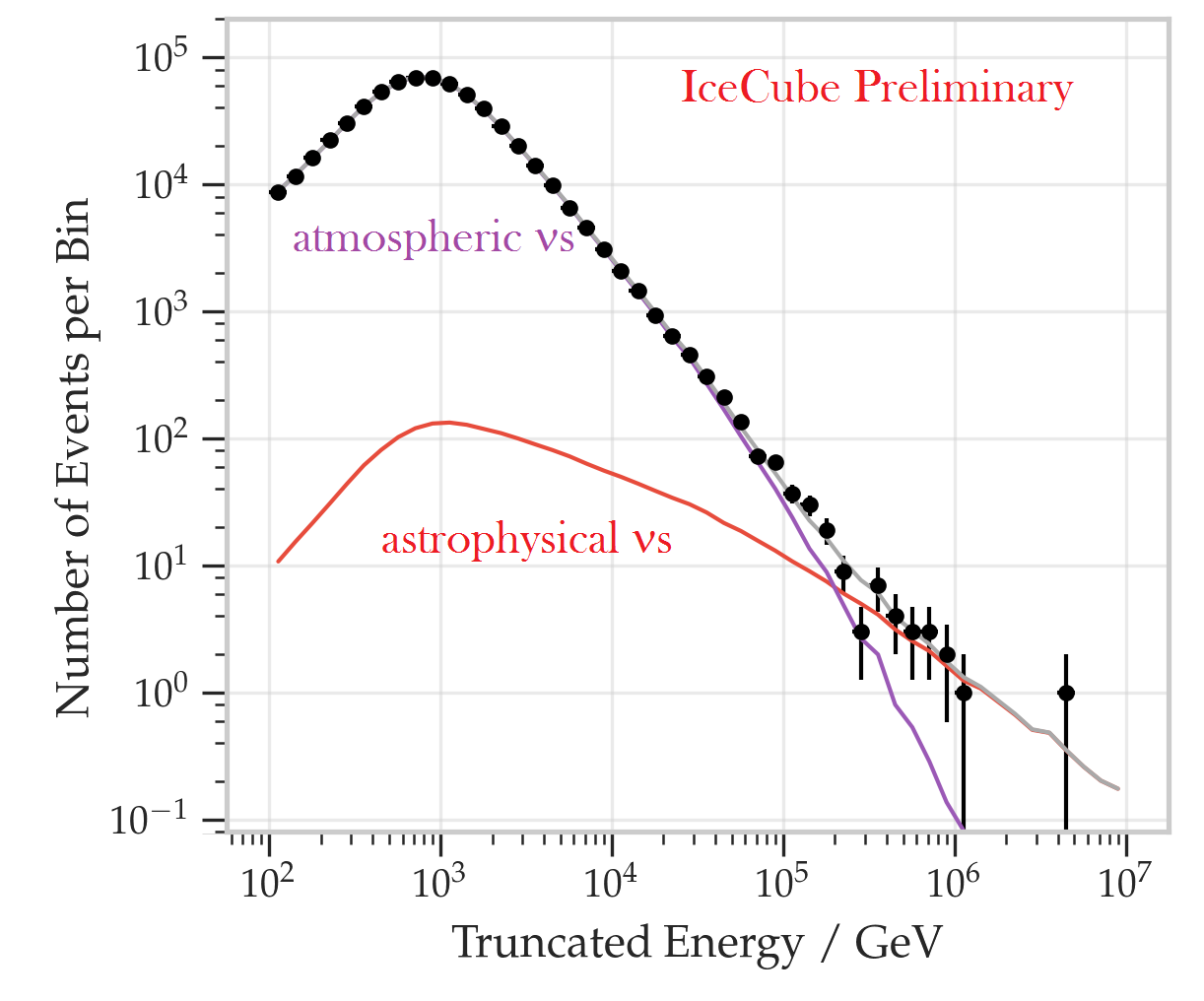}
\end{center}
\caption{\label{fig:10ICpass}
\small 
Unfolded $\nu_\mu$ event distribution from 
 the passing muon sample in IceCube. The~data (crosses) are compared to the best-fit fluxes for atmospheric and astrophysical neutrinos. From~the arXiv version of~\cite{IC1017}.} 
\end{figure}

For these events, the~reconstructed energy $ E^i_{rec}$ of each individual neutrino $i$ is a poor proxy of the true neutrino energy, $E_\nu$. 
Thus, the~reconstructed neutrino energy is used to produce a response matrix $P(E^i_{rec}; E_\nu)$, which must be inverted to produce the posterior probability density function $P(E_\nu; E^i_{rec})$ \cite{6yr}.
Figure~\ref{fig:10ICpass} shows the distribution of the observable energy for the eight-year sample. 
A clear excess above $\sim$ 100 TeV is visible and is not compatible with the atmospheric background~expectation.

When the atmospheric neutrino background is removed, the~best fit to the full data-set results in an astrophysical power-law flux for one neutrino flavor (i.e., $\nu_\mu+ \bar\nu_\mu$) \cite{IC1017}:
\begin{equation}\label{eq:10pass}
\frac{d \Phi_{2\nu}}{dE}= (1.41^{+0.25}_{-0.24})\cdot 10^{-18} 
\biggl( \frac{E}{100 \textrm{ TeV}} \biggr)^{-(2.28\pm{0.09})} \textrm{ GeV}^{-1}\textrm{ cm}^{-2}
\textrm{ s}^{-1}\textrm{ sr}^{-1}  \ .
\end{equation}

\subsection{The Observation from TXS 0506 + 056}\label{sez:txs}
\label{sec:txs}

On 22 September 2017 (the month following the first observation of the coalescence of the two neutron stars with gravitational waves), the~IceCube collaboration detected a track event induced by a $\sim$300 TeV $\nu_\mu$. The~detection generated an automatic alert that caused related searches from the direction of the event by many experiments. 
The Fermi-LAT satellite telescope reported that the direction of the neutrino was coincident with a known gamma-ray source, the~active galaxy known as TXS 0506 + 056 (object classified as a blazar), that was in a particularly active state at the time of the neutrino detection. 
In addition, the~MAGIC gamma ray telescope (located at the Canary Islands) also observed a significant photon flux of energies up to 400 GeV from the direction of the blazar;  
{the study of the astronomical object was completed by observations at other wavelengths: radio, optical and X-ray.
These observations are compatible with the position of a known blazar at redshift $z= 0.3365 \pm0.0010$ \cite{paiano}.}
Neutrino correlation with the registered activity of TXS 0506 + 056 was classified as statistically significant at the level of 3 standard deviations~\cite{txmul}.

Triggered by this association, the~IceCube collaboration analyzed their data (corresponding to 9.5 years) to search for a neutrino emission in excess to the background from the direction of the blazar. 
An excess was found in the period between September 2014 and March 2015, with~statistical significance estimated at 3.5 standard deviations, independent of and before the 2017 flaring episode~\cite{txic}.

The reconstructed direction of the IceCube neutrino corresponded, at~the location of the ANTARES detector, to~a direction 14.2$^\circ$ below the horizon. A~possible neutrino candidate would thus be detected as an up-going event. A~time-integrated study performed by the collaboration over a period from 2007 to 2017 fits 1.03 signal events, which corresponds to a probability that the background simulates this signal of 3.4\% (not considering trial factors) \cite{antaTXS}.

\subsection{Studies of the Galactic~Region}\label{sez:anta}

Due to their relative proximity, the~possibility of studying Galactic sources is particularly~intriguing.

\begin{figure}[t]
\begin{center}
\includegraphics[width=0.75\textwidth]{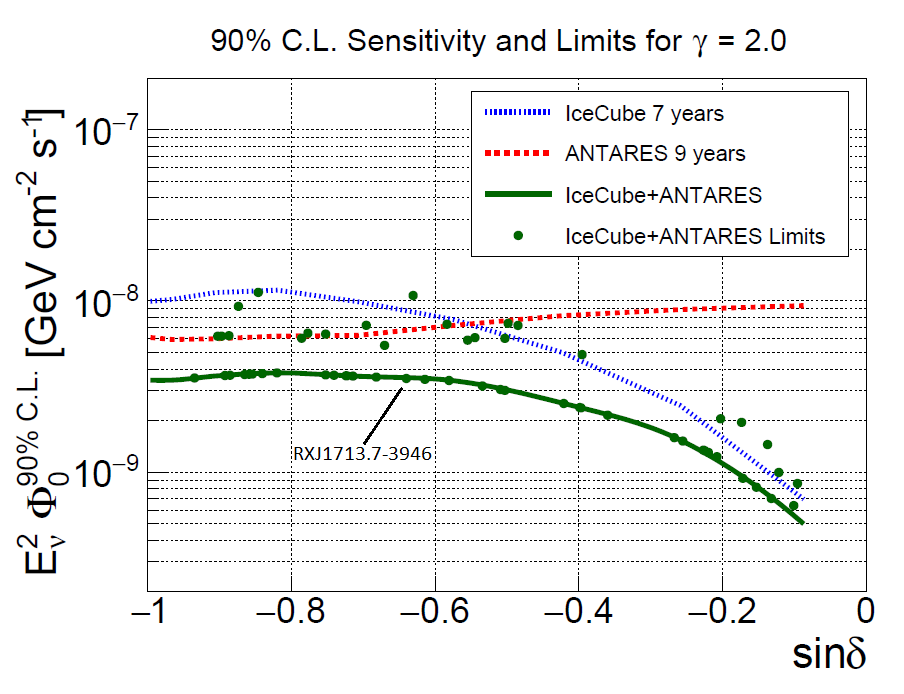}
\end{center}
\caption{\label{fig:SouthSky}
 \small Upper limits at 
 90\% C.L. on the $\nu_\mu+\bar\nu_\mu$ neutrino flux normalization from 57 neutrino candidate sources (green dots) studied by the ANTARES and IceCube collaborations, as~a function of the source declination~\cite{ANICPS}. Such kind of plots depends strongly on the assumption of the neutrino energy spectrum and on the differential sensitivity of the detectors. The~present plot is computed assuming an unbroken neutrino spectrum, with~ spectral index $\gamma=2.0$. For~this spectrum, most of signal is due to neutrinos with $E_\nu>10$ TeV. 
The green line indicates the sensitivity of the combined analysis. The~dashed curves indicate the sensitivities for the IceCube (blue) and ANTARES (red) individual analyses. 
Referring to Figure~\ref{PSkm3}, the~limit corresponding to the SNR RXJ1713.7-3946 is indicated (when comparing the two plots, remember that the y-axis of this figure must be multiplied by $10^{-3}$ to have the same units). }
\end{figure}

The ANTARES detector in the Northern hemisphere can measure \textit{upgoing events}, exploiting, with~respect to IceCube, its complementing field-of-view, exposure, and~lower energy threshold~\cite{ANTAps}. 
In particular, ANTARES is sensitive at a level compatible with the larger IceCube detector to possible sources located in the Galactic plane.
Although a sizable neutrino flux from individual sources is expected to be observable only with a km$^3$-scale detector (see Figure~\ref{PSkm3}), detailed searches for point-like and extended sources of cosmic neutrinos from the Galactic region  have recently been performed using data collected by the ANTARES and IceCube neutrino telescopes~\cite{ANICPS}. 
They combined all the track-like and shower-like events pointing in the direction of the Southern Sky included in a previous nine-year ANTARES point-source analysis, with~the through-going track-like events used in a seven-year IceCube point-source search. The~advantageous field of view of ANTARES and the large size of IceCube are exploited to improve the sensitivity in the Southern Sky by a factor of $\sim$2 compared to both individual analyses.  A~special focus was given to the region around the Galactic Centre, whereby a dedicated search at the location of SgrA* was performed, and~to the location of the supernova remnant RXJ 1713.7-3946. No significant evidence for cosmic neutrino sources was found; upper limits on the flux are presented in Figure~\ref{fig:SouthSky}.

In  this figure, the~upper limit corresponding to the SNR RXJ1713.7-3946 is highlighted in order compare it with the predictions reported in Figure~\ref{PSkm3}. The~predicted flux at $E_\nu>10$ TeV is of the order of few$\times 10^{-9}$ GeV cm$^{-2}$ s$^{-1}$, which is still slightly below the combined ANTARES (9 years) and IceCube (7 years) sensitivities.

\begin{figure}[t]
\begin{center}
\includegraphics[width=0.6\textwidth]{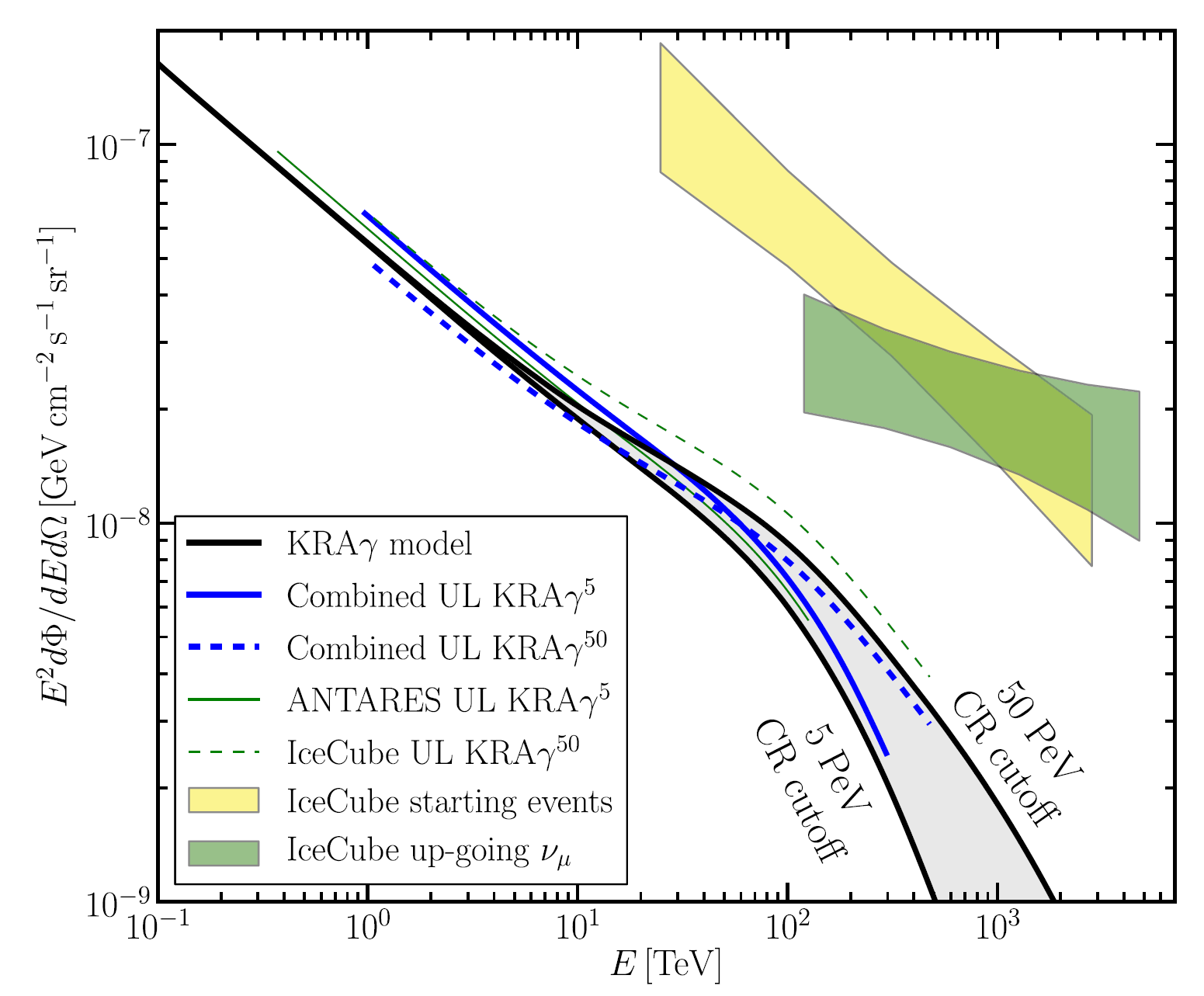}
\end{center}
\caption{\label{fig:galplane}
\small 
Combined upper limits
 from ANTARES and IceCube (at 90\% confidence level, blue lines) on the three-flavor neutrino flux of the model of Galactic cosmic ray propagation (the black lines shows two values of the knee position). The~boxes represent the diffuse astrophysical neutrino fluxes measured by IceCube using an isotropic flux template with starting events (yellow) and upgoing tracks (green) \cite{ANIC}. }
\end{figure}

In addition to individual sources, the~existence of a diffuse Galactic neutrino production is expected from cosmic-ray interactions with Galactic gas and radiation fields. These interactions are also the dominant production mechanism of the diffuse high-energy $\gamma$-rays in the Galactic plane, which have been measured by the Fermi-LAT.
Different models propagating charged cosmic rays diffusively in the interstellar medium producing $\gamma$-rays and neutrinos via interactions with the interstellar radiation field and interstellar gas have been developed. The~ANTARES and IceCube combined their analyses to test~\cite{ANIC} a specific neutrino production model (Figure \ref{fig:galplane}).
As a result, their estimate yields a non-zero diffuse Galactic neutrino flux for a $p$-value of 29\% for the most conservative~model.  

The result of this analysis limits the total flux contribution of diffuse Galactic neutrino emission to the total astrophysical signal reported by IceCube above 60 TeV to be less than 10\%.

\section{What Are the Main Sources of Cosmic Neutrinos?}\label{sec:CRsignal}

The diffuse extraterrestrial flux of high-energy neutrinos, highlighted by IceCube and discussed above, has characteristics that suggest significant extragalactic contributions.
This seems to be confirmed by the observation concerning TXS 0506 + 056.
In spite of that particular association, the~individual sources of IceCube neutrino events remain unidentified~\cite{muretto}. 
 
At present, the~IceCube and ANTARES collaborations manage a series of alert and follow-up programs, which react in real time to events due to neutrino interactions classified as particularly interesting. In~this case, a~short message is automatically sent to the {Gamma-ray burst Coordinates Network (the GCN network) \cite{gcn}}, which activates the radio, optical, X-ray and gamma telescopes. Conversely, particular transient phenomena observed with other probes can be immediately studied with the network of neutrino~telescopes. 

Although we do not have yet a clear theoretical idea on which the sources of cosmic neutrinos are and how intense they are, there are several ideas on the subject. In~this section, largely based on Reference~\cite{epj+}, we discuss a few cases when cosmic neutrinos are directly connected to the (sources of the) cosmic~rays. 

\subsection{Cosmic Rays and Cosmic~Neutrinos} 
Consider an astrophysical site where cosmic rays are produced and confined for some time, 
{and suppose that the same region also contains a `thin' target---} 
much more diffuse than in Earth's atmosphere. 
As described in Section~\ref{sec:CRng}, we might imagine that this site will be particularly brilliant in neutrinos and $\gamma$-rays, and~that the shape of these secondary particles will reflect the production spectra of cosmic rays. 
The cosmic ray spectrum in the source is unknown {\em a priori}, however, general theoretical arguments are in support of the idea that the spectrum behaves as $\sim E_{\mathrm{CR}}^{-2}$, both for $\gamma$-rays and for neutrinos.
(1)~First, such a spectrum is expected for primary cosmic rays in the Fermi acceleration mechanism. 
(2)~Second, if~the target for the cosmic ray collisions is composed by protons or other nuclei, but--differently from the atmosphere--the target layer is thin and there is no significant absorption of the mesons, the~spectra of the secondary particles reflect closely the shape of the primary cosmic ray spectrum. 
(3)~Third, it is not plausible that neutrinos suffer absorption in the source - while for gamma rays, this is possible. In~these conditions, the~spectrum of the secondary neutrinos (and possibly of the associated $\gamma$-rays) till some maximum energy, is expected to be distributed as 
$ \Phi_\nu \propto E_{\nu}^{-2}$.
This spectrum would stand out over the atmospheric neutrino background at sufficiently high energies - see again Figure~\ref{fig-n}.

\subsection{Extragalactic~Sources}
{Nowadays, extragalactic} {sources are believed to give the dominant contribution to the high energy neutrino flux.} 
Assuming that the highest energy cosmic rays that we observe on Earth are typical of the entire cosmos and are of extragalactic origin, we estimate that they have an energy density of $\rho_{\mathrm{uhecr}}=3\times 10^{-19}$ erg/cm$^3$ above 1 EeV. 
Considering the typical evolution time of $T_H=10$ billion years, the~corresponding energy losses of the universe are $W=\rho_{\mathrm{uhecr}}/T_H=9\times 10^{44}\,\mathrm{erg/(Mpc}^3\,\mathrm{yr)}$.
This cosmic ray population will be 
 in equilibrium if, in~the reference volume of 1 Gpc$^3$, there is a population of 
 900 $\gamma$-ray burst{s} and each one injects suddenly $10^{51}$ erg in cosmic rays; {an alternative would be that} there is a population of 150 active galactic nuclei, and~each one radiates continuously $2 \times 10^{44}$ erg/s.
Interestingly, in~both cases, the~number of sources is reasonable and the presumed amount of energy emitted in cosmic rays corresponds to the visible electromagnetic~output.

There are many potential astrophysical sources of high energy neutrinos and below we provide a brief summary of the most significant ones, examining AGNs, blazars, starburst galaxies and Gamma-Ray Bursts (GRBs).

{
\subsubsection{Active Galactic~Nuclei}
An active galactic nucleus (AGN) is a compact region at the center of a galaxy whose luminosity is much higher than the normal one over some portion of the electromagnetic spectrum, with~characteristics indicating that the excess luminosity is not produced by stars. This excess (in the non-stellar emission) has been observed in the radio, microwaves, infrared, optical, ultra-violet, X-ray and gamma ray wavebands. The~radiation from an AGN is believed to be a result of accretion of matter by a supermassive black hole at the center of the host galaxy~\cite{Urry:1995mg}. 
The central engine can accelerate protons up to very high energies, while the accretion disc is an emitter of hot thermal radiation, which gives prominent feature in the observed AGN spectra, usually refereed to as a ``Big Blue Bump''. Accelerated particles move along two jets perpendicular to the accretion disc and crossing this radiation field. 
AGNs are characterized by a strong positive cosmic evolution, meaning that these objects were much more abundant in the~past.

AGN’s are considered potential sites {for high energy neutrino production since 30 years} \cite{Stecker:1991vm,Stecker:2013fxa,Kalashev:2014vya}. High energy neutrino appear in charged pion decays created in proton-$\gamma$ interactions, Equation~(\ref{eq:pg1}), 
due to the collision between protons and the ”blue bump” photons. 
Protons can be accelerated up to $\sim$EeV energies and absorbed in the radiation field contained in the disk, that has a temperature of the order of 10 TeV and a black hole mass of $\sim$$10^8 M_\odot$. Therefore the resulting neutrino flux is expected to be relevant in the sub-PeV and in the PeV region. 
One clarification is needed: as explained in Section~\ref{sec:hm}, neutrinos from $pp$ or $p\gamma$ interactions takes 5\% of the primary proton energy. However, during~the propagation, neutrinos looses energy adiabatically, reaching the Earth with an energy equal to $(E_p /20) /(1+z)$, where $E_p$ is the primary proton energy. Since the distribution of AGNs is strongly positive, the~larger contribution to the neutrino emission is provided by sources having redshift between $z=1$ and $z=2$. It means that most of high energy neutrinos reaching Earth, will have an energy $\sim$50 times lower than the energy of the primary~proton. 

Nowadays it is not possible to exclude (or confirm) the AGNs as potential sources of a part of the IceCube signal. The~increasing of the statistics can help in this game, although~it is hard to constrain certain class of abundant sources (such as low luminosity AGNs).


\subsubsection{Blazars}\label{blazar:sources}
Blazars are AGNs with the emitted jet pointing toward the Earth. They are divided in two classes: BL  Lacertae (BL Lac) and Flat Spectrum Radio Quasar (FSRQ). FSRQs are characterized by the presence of a broad line region and by emission lines in the optical spectrum, while BL Lacs are characterized by featureless optical spectra (i.e., lacking strong emission/absorption lines). A~complete review of the processes occurring in blazars is provided in Reference~\cite{Boettcher:2013wxa}.

The Fermi-LAT satellite proved that the blazars are the brightest extragalactic sources above 10 GeV. Many more of these blazars are not visible as point sources, being too faint and/or too far from us. They contribute to the unresolved $\gamma$-ray radiation, again observed by Fermi-LAT. The~sum of resolved and unresolved blazars almost saturates the observed diffuse emission, providing $\sim 80\%$ of the Extragalactic Gamma-ray Background (EGB) above 100 GeV~\cite{TheFermi-LAT:2015ykq}, with~a margin of uncertainty of some ten percents, that should be attributed to other~sources.

Since blazars dominate the $\gamma$ emission, it is natural to consider them also as high energy neutrino emitters~\cite{Keivani:2018rnh,Murase:2018iyl,Palladino:2018lov,Kadler:2016ygj,no1,Murase:2014foa}.
Particularly, FSRQs are expected to be very efficient in the neutrino production, due to the interaction that can occur in the broad line region. Moreover, as~discussed in Section~\ref{sec:txs}, up~to now the only confirmed sources of one high energy neutrino is a~blazar. 

However FSRQs (and high luminosity blazars in general) are not abundant in the universe. To~have an idea, blazars having a gamma-ray luminosity larger than $L_\gamma = 10^{45} \mbox{ erg/s}$ have a local density of 10 sources per Gpc$^{-3}$. For~these reasons if high luminosity blazars were the sources, at~least two or more neutrinos from the one FSRQ would be expected. This is not observed in the present neutrino data, meaning that blazars cannot contribute more than 20\% to the observed high energy neutrino flux~\cite{no1}. It is important to remark that this result does not say anything on the possible contribution of blazars at higher energies. For~example, blazars may dominate the emission in the multi PeV-EeV energy range, but~nowadays we do not have the technology to observe such a flux, since the flux decreases with the increasing of the energy in any plausible production~mechanism. 

In Figure~\ref{fig:multiplet} the problem due to the absence of multiplets is clearly shown. 
{A multiplet is defined as two or more events from the same source (within the angular resolution of track-like events, order of 1$^\circ$) during the time window in which data have been collected by IceCube. Up~to now no multiplets are present in the IceCube measurements.}
The purple region represents the local power density required to match the flux measured by IceCube. The~blue regions denote the parameter space excluded from the absence of multiplets in the neutrino data. We see in the left panel that sources that are brilliant but rare (like FSRQs) are excluded by the present~measurements. 

\begin{figure}[t]
\centering
\includegraphics[scale=0.28]{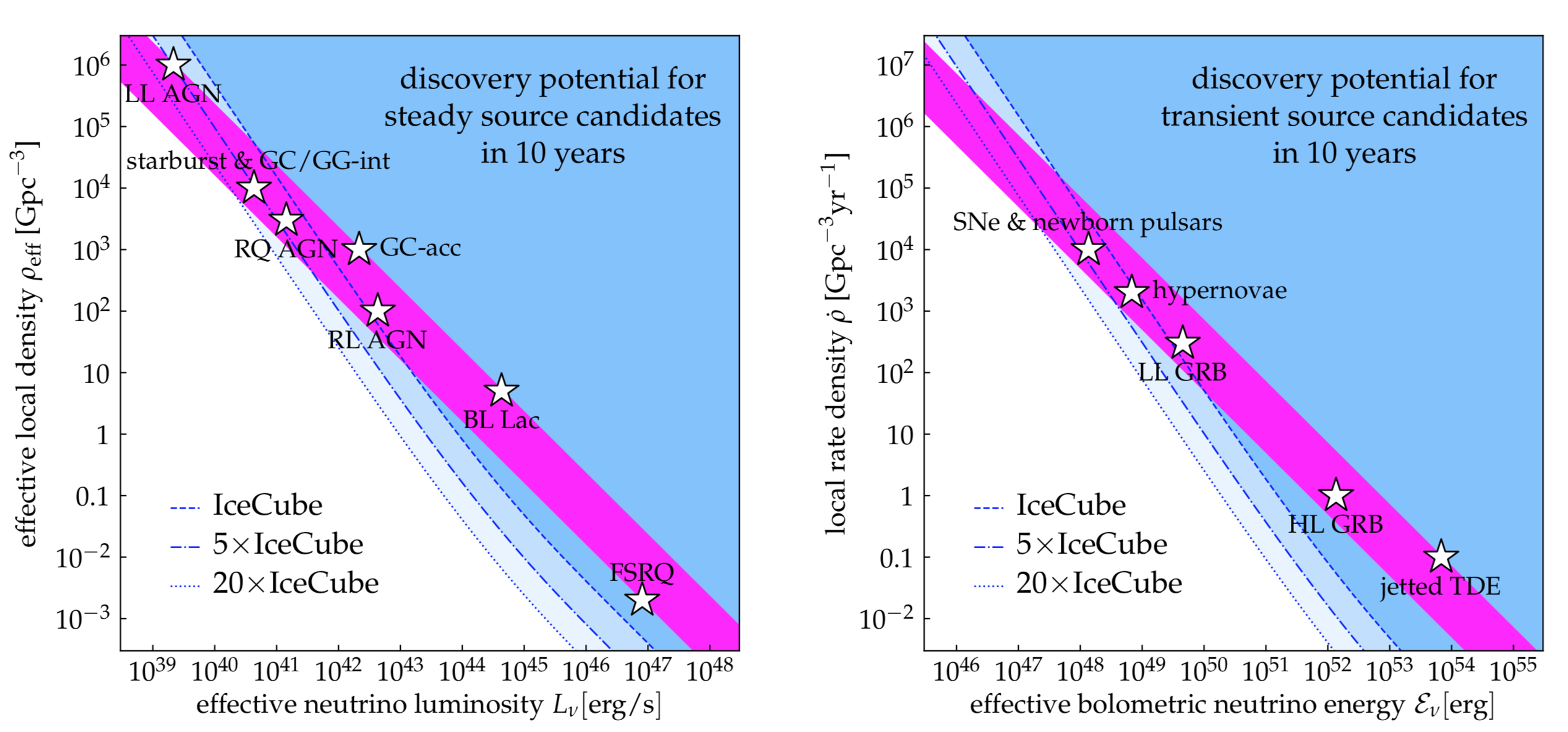}
\caption{\small Figure from Reference 
~\cite{Ackermann:2019ows}. The~purple bands show the power required to interpret the flux of astrophysical neutrinos detected by IceCube (steady sources on the left, transient sources on the right). The~blue regions show the parameter space excluded by the absence of multiplets  in the neutrino data (i.e., two or more events from the same source). Sources that are abundant and faint (top left of the panels) are favorite by the data.}
\label{fig:multiplet}
\end{figure}
\unskip

\subsubsection{Starburst~Galaxy} \label{starb:sources}
The starburst galaxies are a subset of star forming galaxies (SFGs) that undergo an episode of vigorous star formation in their central regions. The~gas density is much higher than what is observed in quiescent galaxies and for this reason the $pp$ interaction is a plausible mechanism to produce high energy neutrinos. Particularly the star formation rate can be 10--100 times higher that the one of Milky Way. Diffusion in starburst galaxies might also become weaker due to strong magnetic turbulence, while advective processes might be enhanced. Since the losses by inelastic collisions and advection are nearly independent of energy, the~hadronic emission of starburst is expected to follow more closely the injected cosmic ray  nucleon spectrum, $E^{-\alpha}$, with~$\alpha \simeq 2.15$ \cite{Loeb:2006tw,Tamborra:2014xia,Chang:2014hua}.

The nearest starburst galaxies are M82 and NGC 253, both at a distance of 3.5 Mpc. These galaxies exhibit relatively hard $\gamma$-ray spectrum in the GeV to TeV energy range, with~a spectral index between 2.1 and 2.3. Especially the $\gamma$-ray spectrum of NGC 253 has been well measured by both Fermi-LAT and H.E.S.S. experiments. Due to the harder emission spectrum and a higher pion production efficiency, the~starburst subset is predicted to dominate the total diffuse $\gamma$-ray emission of SFGs beyond a few GeV, while above 10 GeV the contribution of blazars becomes dominant. 
Provided that the cosmic ray accelerators in starburst galaxies are capable of reaching energies of $\sim 10$ PeV per nucleon, the~hadronic emission can also contribute significantly to the diffuse neutrino emission at PeV energies. However a problem related to the Multimessenger connections between neutrinos and $\gamma$-rays comes out in this~scenario. 

Indeed, in~proton-proton ($pp$) interaction an about equal amount of $\pi^+, \pi^0$ and $\pi^-$ is produced. 
 Taking into account also the sub-dominant contribution of kaons, the~emitted all flavor neutrino flux is almost equal to the emitted $\gamma$-ray flux. After~the propagation the two fluxes become very different, because~neutrinos lose energy only adiabatically while $\gamma$-rays at TeV scale interact with the Extragalactic Background Light (EBL) and the energy is redistributed in the 1--100 GeV energy range approximately with a $E^{-2}$ spectrum. However, it is possible to connect the measured neutrino flux with the associated expected $\gamma$-ray, taking into account the effect of the propagation. This has been done in two works, finding different results. If~one relies on the IceCube HESE sample (Section \ref{sez:hese2}), the~measured neutrino spectrum is very soft and the associated gamma-ray flux would be too high~\cite{Bechtol:2015uqb}, overshooting the possible contribution to the extragalactic background expected from Starburst Galaxies ($\sim$ 20\% above 50 GeV). Following this result, the~contribution of starburst galaxies would be at level of 10\% (see left panel of Figure~\ref{fig:multim}). However if one relies  on the through-going muon flux, starburst galaxies are still perfectly compatible with astrophysical neutrinos~\cite{Palladino:2018bqf} (see right panel of Figure~\ref{fig:multim}). Moreover the absence of multiplets in neutrino data suggests that high energy neutrinos are produced by abundant and faint sources; following this argument, starburst galaxies would be an ideal candidate. Future detections are required to improve the knowledge of the origin of astrophysical neutrinos. }
\begin{figure}[t]
\centering
\includegraphics[scale=0.23]{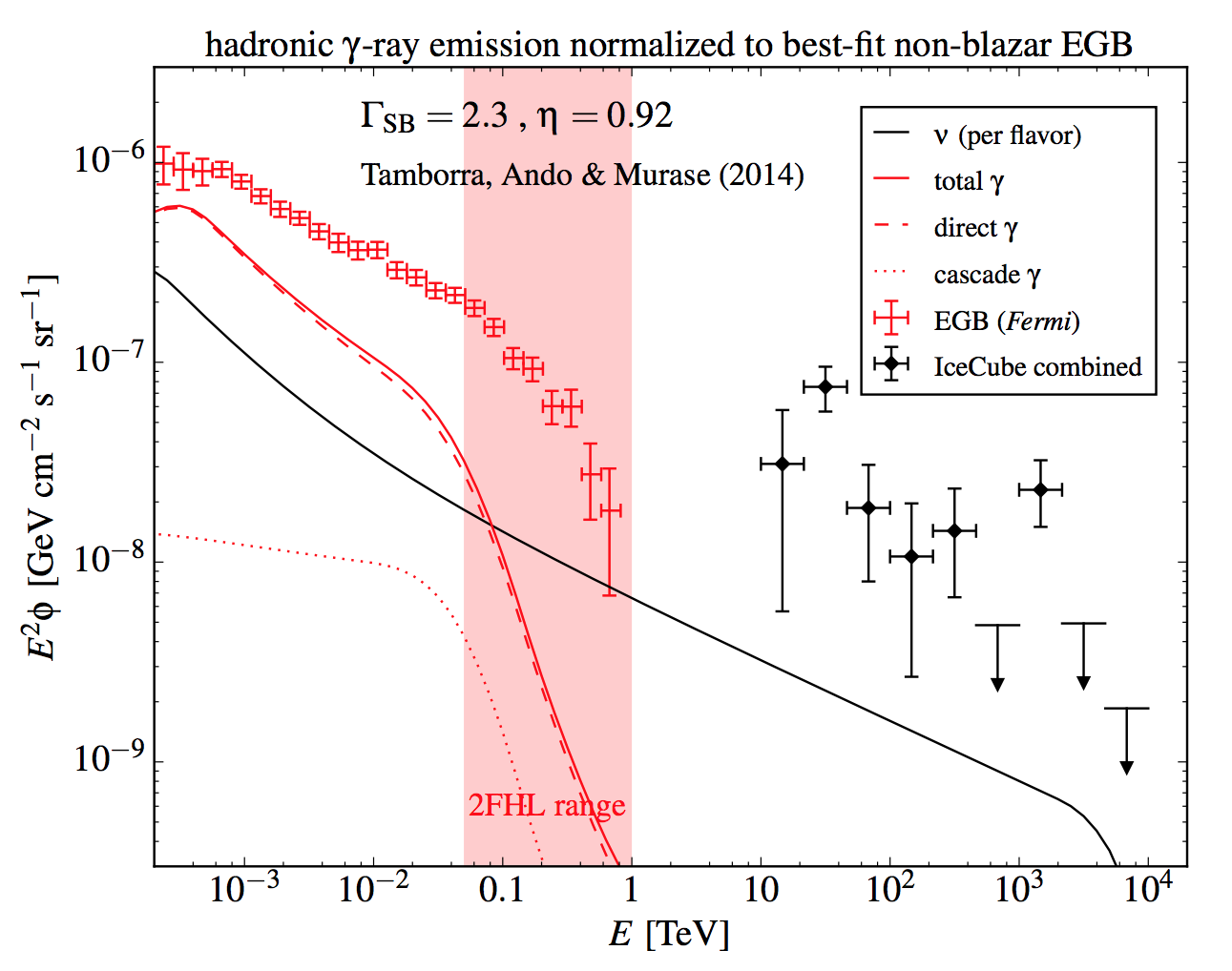}
\includegraphics[scale=0.3]{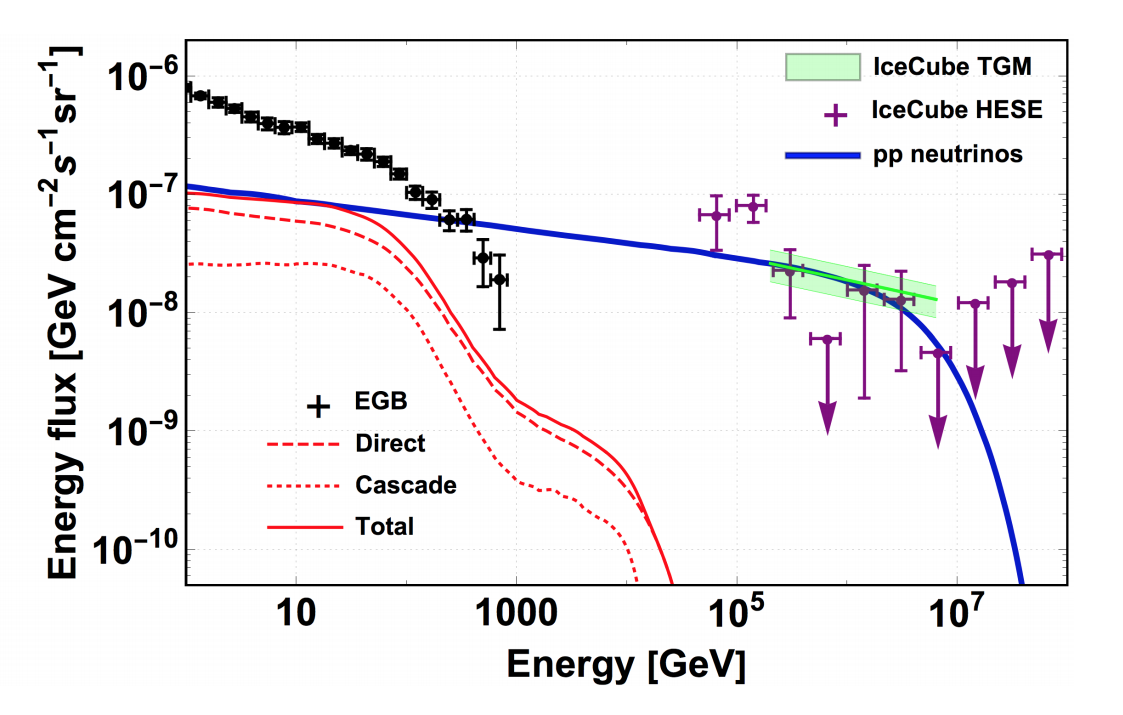}
\caption{\small Multimessenger comparison
 between the flux of astrophysical neutrinos and the associated flux of $\gamma$-rays. In~the left panel, High Energy Starting Events for the neutrino flux are used (figure from Reference~\cite{Bechtol:2015uqb}). In~the right panel through-going muons are used (figure from Reference~\cite{Palladino:2018bqf}). }
\label{fig:multim}
\end{figure}

\section{Discussion and~Prospects\label{sec:conc}}
The present status of neutrino astrophysics is extremely exciting, with~very intriguing perspectives. 
The first general remark is that the total number of cosmic neutrinos is small. Due to the not null background from atmospheric neutrinos, usually each neutrino candidate is tagged with an \textit{event signalness}. This  is a parameter (used firstly by the IceCube collaboration to classify their neutrino candidates) defined as the ratio of the astrophysical expectation over the sum of the atmospheric and astrophysical expectations for a given energy proxy and a specific neutrino flux. Events with high signalness ($>$0.5) are usually high-energy events. {The signalness is dependent on assumptions, for~example, on~the cosmic neutrino flux that is derived from data itself.}
The number of events with high signalness in IceCube are about 3 passing-events above 200 TeV a year (where 1 of them on average should be due to background) and about twice as many HESE events with deposited energy $> 60$ TeV. 
The features of the energy flux for the two samples are in good agreement at high energy.
The plausible interpretation of the flux supports strongly the view that extragalactic high-energy cosmic neutrinos have been observed.  This emission is \textit{diffuse} emission: probably, this diffuse emission is, at~least in part, due simply to \textit{still unresolved} emission.

Although smaller than IceCube, the~ANTARES detector studied the diffuse emission from the Southern sky using both track- and shower-like events. They found, using data collected in nine years, a~mild excess of high-energy events over the expected background; the excess is compatible with the IceCube findings~\cite{ANTAdiff}. 

The distribution that fits the data-set of passing-events collected, Equation~(\ref{eq:10pass}), fits also the high energy subset of the HESE, Equation~(\ref{eq:10hese}); however the low energy part of the HESE does not agree with Equation~(\ref{eq:10pass}).  
In order to explain the discrepancy, the~hypothesis of an additional component present in the HESE, but~not in the passing-event data-set, has been considered
~\cite{spuprd,chen,palvis,vince}.
If (most of) the  cosmic neutrinos have an extragalactic origin, an~isotropic distribution is to be expected. This is 
in first approximation in agreement with the observations of IceCube. 
A contribution due to a Milky Way emission cannot be excluded, and~a value of the order of $\sim 10$\% at $E_\nu>60$ TeV is not incompatible with observations~\cite{ANIC}. 

The identification of sources of the high energy neutrinos relies on the possible correlation of observed events with  specific objects in the sky and/or by self-correlating the events themselves. 
A part with the relevant exception of the coincidence between the direction of a very-high energy $\nu_\mu$ and the position of a blazar, as~discussed in Section~\ref{sez:txs}, searches of correlations among neutrino candidates and classes of astrophysical  objects contained in gamma-ray catalogs  did not reveal a significant amount of coincidences~\cite{no1,no2,no3}. 

From theoretical models, the~timescale of the neutrino emission depends dramatically upon the type of sources.
The hypothesis that the gamma ray burst are associated to a prompt neutrino emission (lasting a few hours or less) has been tested:
{the non-detection of coincident events between neutrinos and GRBs with IceCube~\cite{ICgrb1,ICgrb2} and ANTARES~\cite{ANTAgrb} data has led to important constraints on cosmic-ray acceleration in GRBs.}
Other episodic events of intense neutrino emission could at least in principle occur in AGN, and~be associated, for~example, also to accretion phenomena. On~the contrary, it is unlikely that stationary populations of cosmic rays (as, say, the~ones that presumably exist in a starburst galaxy) may lead to this type of~phenomenology.

The current status of experimental results is not inconsistent with expectations, namely that there exist astronomical sites where cosmic rays distributed as $E_{\mbox{\tiny CR}}^{-2}$ collide with surrounding hadrons leading to intense neutrino fluxes, whose distribution reflects closely the cosmic ray spectra. 
However, it is quite evident the interest in probing accurately the shape and the extent of the observed spectrum, checking it at very high energies, observing events at the Glashow resonance energy, tau~neutrinos.

A key role will be also the observation of cosmic events below few tens of TeV against the background of atmospheric events, using the improved angular resolution of incoming detectors. 
In the quest for the identification of  (some of) the sources, the~muon events (that induced by $\nu_\mu$ CC interactions) have the best angular resolution and offer the best chances.
The improved angular resolution is one one of the main motivation for large volume neutrino telescope in water, as~foreseen by the KM3NeT project. 
 Other reasons are the need to verify the current findings/discoveries and the wish to explore new patches of the sky - including the central regions of the Milky Way. 
Last but not least, when reliable observational information will be available,  it will be possibly to study interesting and possibly important sites for cosmic ray production and the environment where they interact.   
For this, the~role of Multimessenger astronomy is of paramount~importance.


\section*{Acknowledgments}

This work was partially supported by the research grant number 2017W4HA7S ``NAT-NET: Neutrino and Astroparticle Theory Network'' under the program PRIN 2017 funded by the Italian Ministero dell'Istruzione, dell'Universit\`a e della Ricerca (MIUR); it has received funding from the European Research Council (ERC) under the European Union's Horizon 2020 research and innovation programme (Grant No. 646623). Figures~\ref{fig:convastro} and \ref{fig:multim}
are reproduced by permission of IOP Publishing:  \copyright IOP Publishing Ltd and Sissa Medialab,  
all rights reserved. 

We are grateful to three anonymous reviewers for the excellent feedback we received.
F.V.~thanks 
F.~Aharonian, 
A.~Capone, 
S.~Celli, 
M.L.~Costantini, 
E.~Esmaili, 
A.~Gallo Rosso, 
P.L.~Ghia, 
C.~Mascaretti,
E.~Roulet and 
V.~Zema for precious~discussions.

\newpage
\appendix

\section{How to Estimate Neutrinos from~Gamma-Rays}\label{appa}
After the discussion and the caveats, we give various practical recipes to connect high energy gamma rays and neutrinos:
In order, to~be definite we focus on $pp$ collisions and quantify the connection between 
neutrinos and gamma rays in various levels of~approximation.

$(i)$ The simplest procedure is just to recall that the leading pion carries about 1/5 of the initial proton, and~that the energy partition in its subsequent decay leads to a similar sharing of energy. Therefore, 
each neutrino carries about 1/20 of the initial proton energy whereas the gamma rays (from  neutral pion) have twice this~energy.

$(ii)$ A very practical recipe, that can be adopted in the case when the cosmic ray spectra obey the exponential cutoff distribution, is the following one~\cite{kappes}
$$
\frac{dN_p}{dE}= A_p E^{-\alpha} \exp\left(-\frac{E}{\epsilon_p}\right)
\Rightarrow
\frac{dN_{\gamma/\nu}}{dE}= A_{\gamma/\nu} E^{-\Gamma_{\gamma/\nu}} \exp\left(-\sqrt{ \frac{E}{\epsilon_{\gamma/\nu}}}\right)
$$
where
$$
A_\nu\approx (0.71-0.16\alpha)  A_\gamma, \ \ 
\Gamma_\nu\approx \Gamma_\gamma\approx \alpha-0,1,\ \ 
\epsilon_\nu\approx 0.59 \epsilon_\gamma\approx \epsilon_p/40
$$

In this approximation, all neutrinos have the same spectrum, mostly due to~oscillations.

$(iii)$ Finally an accurate approximation allows us to find the neutrino flux $F_\nu$ from the gamma ray flux $F_\gamma$ in the same assumptions,  simply by performing one integral with a known kernel that accounts for the kinematics of the decay 
as shown in Reference~\cite{pat}, based on the results of Reference~\cite{volkova}, and~as further improved in Reference~\cite{vils}. 
For example, the~sum of the muon neutrinos and antineutrinos is given by
$$
\Phi_{\nu_\mu}(E_\nu)+\Phi_{\bar\nu_\mu}(E_\nu)= 
0.66\, \Phi_{\gamma}\left(\frac{E_\nu}{1-r_\pi}\right) + 
0.02\, \Phi_{\gamma}\left(\frac{E_\nu}{1-r_K}\right) + 
\int_0^1 \kappa(x)\;  \Phi_{\gamma}\left(\frac{E_\nu}{x}\right)  dx
$$
where the first term comes from pion decay ($r_\pi=(m_\mu/m_\pi)^2$), the~second from kaon decay
($r_K=(m_\mu/m_K)^2$), the~third from the muons, and~$$
\kappa(x)=\left\{
\begin{array}{ll}
x^2 (33.8-54.3 x),& \mbox{ if } x<r_K \\
(1-x)^2 (-0.63 + 12.45 x),& \mbox{ if } x>r_\pi \\
0.04+0.20 x+7.44 x^2-7.53 x^3& \mbox{ otherwise } \\
\end{array}
\right.
$$ 

See Reference~\cite{prompo} for the other flavors and for the 
most updated expression of the kernels.  
Note that in the last method, the~assumption of a power law distribution is not necessary.
Of course, in~all of these cases the gamma ray flux is supposed to be fully due to  `hadronic' origin and to be unperturbed from propagation, as~discussed just above.

%
%





\newpage
\parskip0.12ex
\small
\tableofcontents

\end{document}